\documentclass{article}
\usepackage{amsfonts, amsmath}
\usepackage{graphicx}
\usepackage{subfig} 
\newcommand{\Z}{\mathbb{Z}}                            
\newcommand{\N}{\mathbb{N}}                            
\newcommand{\R}{\mathbb{R}}                            

\newcommand{\set}[1]{\{#1\}}                           
\newcommand{\normtwo}[1]{\lVert #1 \rVert}             
\newcommand{\normH}[1]{\lVert #1 \rVert _\mathcal{H}}  
\newcommand{\innH}[1]{<#1>_\mathcal{H}}                
\title{Clustering functional data using wavelets}
\author{  Anestis Antoniadis\thanks{
             Universit\'e Joseph Fourier, Laboratoire LJK, 
             Tour IRMA, BP53, 38041 Grenoble Cedex 9, France 
             {anestis.antoniadis@imag.fr}}  \and
          Xavier Brossat\thanks{
             EDF R\&D, 1 avenue du G\'en\'eral de Gaulle, 
 	     92141 Clamart Cedex, France  
 	     {xavier.brossat@edf.fr}}         \and
          Jairo Cugliari\thanks{
 	     EDF R\&D; Universit\'e Paris-Sud, France 
 	     {jairo.cugliari@math.u-psud.fr}}  \and
          Jean-Michel Poggi\thanks{ 	
             Universit\'e Paris 5 Descartes;
 	     Universit\'e Paris-Sud, France
 	     {jean-michel.poggi@math.u-psud.fr}}
}
\date{27 December 2010}
\begin{document} 
%
%
\maketitle
\begin{abstract}
We present two methods for detecting patterns and clusters in high dimensional time-dependent functional data. Our methods are based on wavelet-based similarity measures, since wavelets are well suited for identifying highly discriminant local time and scale features.  The multiresolution aspect of the wavelet transform provides a time-scale decomposition of the signals allowing to visualize and to cluster the functional data into homogeneous groups. For each input function, through its empirical orthogonal wavelet transform the first method uses the distribution of energy across scales generate a handy number of features that can be sufficient to still make the signals well distinguishable. Our new similarity measure combined with an efficient feature selection technique in the wavelet domain is then used within more or less classical clustering algorithms to effectively differentiate among high dimensional populations. The second method uses dissimilarity measures between the whole time-scale representations and are based on wavelet-coherence tools. The clustering is then performed using a k-centroid algorithm starting from these dissimilarities. Practical performance of these methods that jointly designs both the feature selection in the wavelet domain and the classification distance is demonstrated through simulations as well as daily profiles of the French electricity power demand.

\end{abstract}
        \section{Introduction}\label{sec:intro}
In different fields of applications, explanatory variables are not standard multivariate observations, but are functions observed either discretely or continuously. Ramsay and Dalzell (1991) gave the name ``functional data analysis'' to the analysis of data of this kind. As evidenced in the work by Ramsay and Silverman (1997, 2002) (see also  Ferraty and Vieu (2006)), a growing interest is notable in investigating the dependence relationships between complex functional data such as curves, spectra, time series or more generally signals. Functional data often arise  from measurements on fine time grids,  and if the sampling grid is sufficiently dense, the resulting data may be viewed as a sample of curves. These curves may vary in shape, both in amplitude and phase. Typical examples involving functional data  can be found when studying the forecasting of electricity consumption, temporal gene expression analysis or ozone concentration in environmental studies to cite only a few. 
\nocite{ramsay1991some}\nocite{ramsay1997functional}\nocite{ramsay2002applied}
\nocite{ferraty2006nonparametric}

Given a sample of curves, an important task is to search for homogeneous subgroups of curves using clustering and classification. Clustering is one of the most frequently  used data mining techniques, which is an unsupervised  learning process for partitioning a data set into sub-groups  so that the instances within a group are similar to each other and are very dissimilar to the instances of other groups. In a functional context clustering helps to identify representative curve patterns and individuals who are very likely involved in the same or similar processes. Recently, several functional clustering methods have been developed such as variants of the k-means method (Tarpey and Kinateder (2003); Tarpey (2007); Cuesta-Albertos and Fraiman (2007)) and clustering after transformation and smoothing (CATS) (Serban and Wasserman (2004)) to  model-based procedures, such as clustering sparsely sampled functional data (James and Sugar (2003)) or  mixed effects modeling approach 
using B-splines (Luan and Li (2003)) mostly concentrate on curves exhibiting a regular behavior.
\nocite{tarpey2003clustering}\nocite{tarpey2007linear}\nocite{cuesta2007impartial}
\nocite{serban2004}\nocite{james2003clustering}\nocite{luan2003}

Our interest in time series of curves is motivated by an application in forecasting a functional time series when the most recent curve is observed. This situation arises frequently when a seasonal univariate time series is sliced into consecutive segments, for example days,  and treated as a time series of functions. The idea of forming a functional time series  from a seasonal univariate time series has been introduced by Bosq (1991) \nocite{bosq1991modelization}. Suppose one observes a square integrable continuous time  stochastic process $X=(X(t), t\in\R)$ over the interval $[0,T]$, $T>0$ at a relatively high sampling frequency.  The commonly used approach is to divide the interval $[0,T]$ into sub-intervals $[ (l-1)\delta, l\delta] $, $l=1,\ldots,n$ with $ \delta = T/n$, and to consider the functional-valued discrete time stochastic process $ Z = (Z_i, i\in\N) $, defined by 
    \begin{equation}\label{eq:X2Z}
	Z_i(t) = X( t + (i-1)\delta ), \;\;\;\;\;\;\; 
		      i\in\N \;\;\; \forall t \in [0,\delta)
    \end{equation}
The random functions $Z_i$ thus obtained, while exhibiting a possibly nonstationary behavior within each subinterval, form a functional times series that is usually assumed to be stationary. Such a  procedure allows to handle seasonal variation of size $\delta$ in a natural way. This set-up has been used for prediction when ones consider a Hilbert-valued autoregressive process (see Bosq (1991); Besse and Cardot (1996); Pumo (1992); Antoniadis and Sapatinas (2003)) or for a more general continuous-time process (see  Antoniadis \textit{et al.}~(2006)). However, as already noticed above, for many functional data the segmentation into subintervals of length $\delta$ may not suffice to make reasonable the stationary hypothesis of the resulting segments. For instance, in modeling the electrical power demand process the seasonal effect of temperature and the calendar configuration strongly affects the mean level and the shape of the daily load demand profile. {\em Recognizing this, our aim is therefore to propose  a clustering technique that clusters the functional times series into groups that may be considered as stationary so that in each group more or less standard functional prediction procedures can be applied.}
\nocite{bosq1991modelization}\nocite{besse1996approximation}
\nocite{pumo1992prediction}\nocite{antoniadis2003wavelet}) 
\nocite{antoniadis2006functional}

We will apply the methodology focusing on \textsc{edf}'s ({\it \'Electricit\'e de France}\footnote{\texttt{http://www.edf.com}}) national power demand for a year. This is essentially a continuous process even though we only count with discrete records sampled at 30 minutes for the whole year. 
Some of the facts associated to the electricity power demand induce to think that the process is not stationary. We will construct a functional data set by splitting  the continuous process as  in (\ref{eq:X2Z}) where the parameter $\delta$ will be a day.

Although slicing an univariate time series produces functional data, we do not observe the whole segments but a sample of the values at some time points. 
One could then use a vector representation of each observation.
For example Wang \textit{et al.}~(2008) proposed to measure the distance between observations through the high dimensional multivariate distribution of all sampled time points along each curve.  This approach does not exploit the eventually potential information of interactions between points of a single curve. To avoid this, many authors have clustered the coefficients of a suitable basis representation of functions. Since the analyzed curves are infinite-dimensional and temporal-structured, one projects each curve over an appropriate  basis of the functional space to extract specific features form the data which are then used as inputs for clustering or classification. One may cite for example Abraham \textit{et al.}~(2003) where the authors proposed to cluster the spline coefficients of the curves using $k-$means, or James and Sugar (2003) that use
a spline decomposition specially adapted for sparsely sample functional data. Nevertheless, attention must be paid to the chosen basis because this operation involves linear transformation of data that may not be invariant for the clustering technique (see Tarpey (2007)). Splines are often used to describe functions with a certain degree of regularity. However, we will be working with curves like in Figure \ref{fig:conso-traj} that may present quite irregular paths. We chose to work with wavelets because of their good approximations properties on sample paths that might be rather irregular.
\nocite{wang2008nonparametric}\nocite{abraham2003unsupervised}
\nocite{james2003clustering}\nocite{tarpey2007linear}

Wavelets offer an excellent framework when data are not stationary. For example, in Gurley \textit{et al.}~(2003) the wavelet transform is used to develop the concept of wavelet-coherence  that describes the local correlation of the time-scale representation of two functions. Grinsted \textit{et al.}~(2004)  proved that this concept is convenient   for clustering geophysical time series. Another example supporting such a fact is the work by Quiroga \textit{et al.} which uses wavelets to detect and cluster spikes on neural activity. Motivated by this and the  fact that the wavelet transform has the property of time-frequency localization of the time series,  we propose hereafter a time-series feature extraction algorithm using orthogonal wavelets for automatically choosing feature dimensionality for clustering. We also study some more complex alternatives using the wavelet-coherence concept in sake of better exploiting the well localized information of the wavelet transform. 
\nocite{gurley2003first}\nocite{grinsted2004application}
\nocite{quiroga2004unsupervised}

The rest of the paper is organized as follows. Section \ref{sec:method}  is a reminder on multiresolution analysis and introduces the basis supporting our feature extraction algorithm by means of the energy operator. Following wavelet analysis we cluster the functional data using the extracted features in Section \ref{sec:kmeans}. Our first clustering algorithm uses $k$-means as  unsupervised learning routine. 
We test the proposed method in Section \ref{sec:sim} on simulated and real data.
Section \ref{sec:cohe} presents a more sophisticated method for clustering functional data using a more specific dissimilarity measure. 
Finally, we conclude the paper by summarizing the main contributions and perspectives in Section \ref{sec:discussion}. 
%
%
%
%
%
%
%
          \section{Feature extraction with wavelets}\label{sec:method}
In this section we first introduce some basic ideas  of the wavelet analysis before introducing more specific material: the energy contributions of the scale levels of the wavelet transform which are the key tools for future clustering. More details about wavelets and wavelet transforms can be found for example in Mallat (1999). \nocite{mallat1999wavelet}
 
We will consider a probability space $(\Omega, \mathcal{F}, P)$ where we define a function-valued random variable $ Z : \Omega \to \mathcal{H} $, where $\mathcal{H}$ is a (real) separable Hilbert space (e.g. $\mathcal{H}=\mathcal{L}^2 (\mathcal{T}) $ the space of squared-integrable functions on $ \mathcal{T} =  [0, 1)$ (finite energy signals) or $\mathcal{H}= {\mathcal{W}^s_2} (\mathcal{T}) $ the Sobolev space of $s$-smooth function on $\mathcal{T}$, with integer regularity index $s\geq 1$) endowed with the Hilbert inner product $\innH{.,.}$ and the Hilbert norm $\normH{.}$.
   \subsection{Wavelet transform} 
A wavelet transform (WT for short) is a domain transform technique for hierarchical decomposing finite energy signals. It allows a real valued function  to be described in terms of an approximation of the original function, plus a set of details that range from coarse to fine. The property of wavelets is that the broad trend of the input function is captured  in the approximation part, while the localized changes are kept in the detail components. For short, a wavelet is a smooth and quickly vanishing oscillating function 
with good localization properties in both frequency and time. This is suitable for approximating  curves that contain localized structures. A compactly supported WT uses a orthonormal basis of waveforms derived from scaling (i.e. dilating or compressing) and translating a compactly supported scaling function $\widetilde\phi$ and a compactly supported mother wavelet $\widetilde\psi$. We consider periodized wavelets in order to work over the interval $[0,1]$, denoting by 

  $$ \phi(t) = \sum_{l\in\Z} \widetilde\phi(t-l) \;\;\;\; \hbox{ and } \;\;\;\;
     \psi(t) = \sum_{l\in\Z} \widetilde\phi(t-l), \;\;\;\; \hbox{ for } 
      \;\;  t \in [0, 1],  $$ 

the periodized scaling function and wavelet, that we dilate or stretch and translate 
  $$ \phi_{j,k}(t) = 2^{j/2} \phi( 2^j t - k ) ,\;\;\;\;
      \psi_{j,k}(t) = 2^{j/2} \phi( 2^j t - k ) . $$ 

For any $j_0 \geq 0$, the collection 
  $$ \set{ \phi_{j_0,k}, k= 0,1,\ldots,2^{j_0}-1; 
            \psi_{j,k},   j \geq j_0, k=0,1,\ldots,2^j-1 },  $$ 
is an orthonormal basis of $\mathcal{H}$. Thus, any function $z \in \mathcal{H}$ can then be decomposed in terms of this orthogonal basis as
  \begin{equation}\label{eq:zeta}
     z(t) = \sum_{k=0}^{2^{j_0}-1} c_{j_0,k} \phi_{j_0,k} (t)  + 
        \sum_{j={j_0}}^{\infty} \sum_{k=0}^{2^j-1} d_{j,k} \psi_{j,k} (t) ,
  \end{equation}
where $c_{j,k}$ and $d_{j,k}$ are called respectively the scale and the wavelet coefficients of $z$ at the position $k$ of the scale $j$ defined as
   $$ c_{j,k} = \innH{ z, \phi_{j,k} }   \;\;\;\;
      d_{j,k} = \innH{ z , \psi_{j,k}  }.  $$
      
To efficiently calculate  the WT, Mallat introduced the notion of mutiresolution analysis of $\mathcal{H}$ (MRA) and designed  a family of fast algorithms (see Mallat (1999)). \nocite{mallat1999wavelet}
 
With MRA, the first term at the right hand side of (\ref{eq:zeta}) can be viewed as a smooth approximation of the function $z$ at a resolution level $j_0$. The second term is the approximation error. It is composed by the aggregation of the details at scales $j \geq j_0$. These two components, approximation and details, can be viewed as a low frequency (smooth) nonstationary part and a component that keeps the time-localized details at higher scales. The distinction between the smooth part and the details is determined by the resolution $j_0$, that is the scale below which the details of a signal cannot be distinguished. We will focus our attention on the finer details,  i.e. on the information at the scales $\set{ j:j \geq j_0 } $.

From a practical of view, each function is usually observed on a fine time sampling grid of size $N$. In the sequel we will be interested in input signals of length $N=2^J$ for some integer $J$. If $N$ is not a power of 2, one may interpolate data  to the nearest $J$ with $2^{J-1} < N < 2^J$. In this context we use a highly efficient pyramidal algorithm (Mallat (1989)) to obtain the coefficients of the Discrete Wavelet Transform (\textsc{dwt}). \nocite{mallat1989theory}

Denote by  $ {\mathbf{z}} = \set{ z(t_l): l=0,\ldots,N_i-1 }$ the finite dimensional sample of the function $z$. For the particular level of granularity given by the size $N$ of the sampling grid, one rewrites (\ref{eq:zeta}) using the truncation imposed by the $2^J$ points and the coarser approximation level $j_0=0$, as:
  \begin{equation}\label{eq:zetaJ}
     \widetilde{z}_J (t)= c_{0}  \phi_{0,0} (t)  + 
              \sum_{j=0}^{J-1} \sum_{k=0}^{2^j-1} d_{j,k}  \psi_{j,k} (t).
  \end{equation}
Hence, for a chosen wavelet $\psi$ and coarse resolution $j_0=0$,  one may define the \textsc{dwt} operator:
    $$W_\psi : \R^N \rightarrow \R^N, \quad \mathbf{z} \mapsto 
        \left(\mathbf{d}_0, \dots, \mathbf{d}_{J-1}, c_0f \right)$$
with $\mathbf{d}_j = \set{d_{j,0}, \ldots, d_{j,2^j - 1}}$.
Since the \textsc{dwt} operator is based on an $L_2$-orthonormal basis decomposition, Parseval's theorem states that the energy of a square integrable signal is preserved under the orthogonal wavelet transform: 
  \begin{equation}\label{eq:energy}  
     \| \mathbf{z} \|_2^2 = c_{0}^2 + \sum_{j=0}^{J-1} \sum_{k=0}^{2^j-1} d_{j,k} ^2  = 
                     c_{0}^2 + \sum_{j=0}^{J-1} \| \mathbf{d}_{j} \|_2^2.
  \end{equation}
Hence, the global energy $\|\mathbf{z}\|_2^2$ of $\mathbf{z}$ is broken down into a few energetic components.  
The way these energies are distributed and contribute to the global energy of a signal is the key fact that we are going to exploit to generate a handy number of features that are going to be used for clustering. 

The image of the \textsc{dwt} operator applied on the column vector $\mathbf{z}$ of dimension $N = 2^J$  may be written in matrix form as:
   $$\mathbf{W}=\mathcal{W}\mathbf{z}$$ 
where $\mathcal{W}$ is a $N$ by $N$ square matrix defining the \textsc{dwt} and satisfying $\mathcal{W}' \mathcal{W} = I_N$ (see \cite[chap. 4]{percival2006wavelet}),
and $\mathbf{W}$ is a column vector of length $N$ with 
   $$ \mathbf{W} = ( \mathbf{d}_0, \mathbf{d}_1, \ldots, \mathbf{d}_{J-1}, c_0 )'$$
where $W'$ denotes the transpose of $W$. It is easy to see that if we consider a vector $\mathbf{x} = a +b\mathbf{z} $ with $a, b \in \mathbb{R}$, then the wavelet coefficients of the \textsc{dwt} of $\mathbf{x}$ are obtained from those of the $\mathbf{z}$ as:
  \begin{equation}\label{eq:transf}
     ( b \mathbf{d}_0, b \mathbf{d}_1, \ldots, b \mathbf{d}_{J-1},  a + b c_0  )' .
  \end{equation}
   \subsection{Absolute and relative contributions} 
We just have seen that \textsc{dwt} coefficients describe properties of functions both at various locations  and at various time granularity. Each time granularity here refers to the level of detail that can be captured by \textsc{dwt}. 
This is therefore the reason of choosing the \textsc{dwt} as a representation scheme in our previous section to compare the shapes of curves for clustering.  
The energy $\mathcal{E}_z =   \normH{ z}^2$ of the time series $z$ via decomposition (\ref{eq:energy}) is equal to the sum of the energy of its wavelet coefficients distributed across scales
   \begin{equation} \label{scaledist}
         \mathcal{E}_z  \approx \normtwo{\mathbf{z} }^2 =
                         c_{0,0}^2 + \sum_{j=0}^{J-1} \normtwo{\mathbf{d}_j }^2,
   \end{equation}
the approximation (denoted informally by $\approx$) holding because of the truncation at scale $J$ for the wavelet expansion of $z$, discarding finer scales. If we consider $\mathbf{z}$ as the difference between two sampled curves, (\ref{scaledist})  justifies using the energy decomposition  of wavelet coefficients for computing squared Euclidean distances between two series. However, when interested to see how the energy of wavelet coefficients is distributed across scales, other distance functions on \textsc{dwt} decompositions may be more appropriate for  measuring the similarity between two series. 

In what follows, we define for $j=0,\ldots,J-1$ the absolute and relative contributions of the scale $j$ to the global energy of the centered function 
\begin{equation}\label{eq:contribs}
     \hbox{cont}_j = \normtwo{ \mathbf{d}_j}^2  \qquad\qquad
     \hbox{rel}_j  = \frac{ \hbox{cont}_j }{ \sum_{j=0}^{J-1} \hbox{cont}_j }
     \qquad  \forall j = 0,\ldots,J-1
  \end{equation}
We call the two different representations: the absolute contribution (AC) and the relative contribution (RC).

We will therefore characterize each time series by the vector of its energy contributions or its relative contributions in order to define an appropriate measure of similarity that is going to be used for clustering.  Note that in both of these choices of representation we leave out the eventual mean level differences of the time series  since  we do not make any use of the approximation term $c_{0,0}$ in their definition. An important consequence of this fact is that any dissimilarity considered over the representations will be invariant under vertical shifts of the curves. Moreover, using RC implies fixing the energy of the curves to one. Hence no difference in amplitude can be detected.
%
%
%
%
%

            \section{A $k-$means like functional clustering procedure}\label{sec:kmeans}
We have presented a way to represent the infinite-dimensional original objects in $J$ features that summarize  the time evolution of the curves at different scales. 
We will now see how we use the information that we have coded to effectively cluster it.

We can sum up the strategy for clustering that we will use in the first part of the paper in the following steps: 
\begin{description}
 \item [\bf{0. Data preprocessing.}] Approximate sample paths of $z_1(t),\ldots,z_n(t)$ by the truncated wavelet series (see (\ref{eq:zetaJ})) at the scale $J$ from sampled data $\mathbf{z}_1, \ldots, \mathbf{z}_n$.
 \item [\bf{1. Feature extraction.}] Compute either the energetic components using absolute contribution (AC) or relative contribution (RC). If using the latter, transform the obtained vector using the logit transformation.
 \item [\bf{2. Feature selection.}] Use a feature selection algorithm for screening irrelevant variables.
 \item [\bf{3. Determine the number of clusters $K$.}] 
 \item [\bf{4. Clustering.}] Obtain by $k$-means algorithm the $K$ clusters using the selected features.
\end{description}

The preprocessing step is mandatory when working with functional data that is only observed on a finite grid. Step 1 extracts from infinite dimensional curves a set of finite dimensional features. This allows us to employ multidimensional data analysis techniques. For the AC we have a vector of positive components that sums up the global energy on the details $\sum_j \normtwo{\mathbf{d}_j}^2_2$. Meanwhile for the RC the vector,  all its positive components sum up one, in other words we have a probability vector. This is a constraint if we want to use the $k-$means algorithm because nothing that warrants that the resulting clusters will be probability vectors. Therefore we transform the vector by using the well known logit transformation.

At this stage we can use the $k-$means algorithm. Before that, a feature selection step is performed. Feature extraction and features selection have really different aims. Whether the former creates some new information from existing objects, the latter only selects a subset of existing features. This selection reduces the computational time of the algorithm and helps to avoid an unsatisfactory and unstable clustering. Another important advantage of using a feature selection algorithm is that the reduced number of features helps to better understand the cluster output. In our case, the number of features  depends on the number of sampling points of the acquired data. For $N$ points, the number of features is $J= \log_2 (N)$ that can be large. Moreover, since we are interested in the energy decomposition across scales, potentially several scales will not be informative for the cluster structure. Besides this, the feature selection algorithm aims to reduce (or eliminate) the presence of nonsignificant variables and a possible redundant information that could hide the cluster structure.

To decide which scales we retain with we use a variable selection algorithm for nonsupervised learning proposed and tested in Steinley and Brusco (2008). This algorithm combines a variable transformation with a variable selection technique. The transformed variables are used to construct an index of clusterability that serves to screen the variables that do not reveal information about the cluster structure. Then, for the remaining variables an evaluation of the feasible subsets of variables is done obtaining the best set of variables (minimum sum-of-square-errors (SSE)) for each subset size. The selected subset of variables is obtained by penalizing the SSE of each subset by its size. 
\nocite{steinley2008new,steinley2008selection}

One of the most difficult task in clustering is the determination of the number of clusters $K$. Even if some statistical support can be given to achieve this task, usually the knowledge on the particular application helps on the choice. In the classical case, i.e. not the functional one, a lot of data-driven strategies can be defined. The first one by inspecting basically the within-cluster dissimilarity as a function of $K$. Many heuristics have been proposed trying to find a ``kink'' in the corresponding plot. 

A more formal argument has been proposed by Tibshirani \textit{et al.}~(2001) by comparing, using the gap statistic, the logarithm of the empirical within-cluster dissimilarity and the corresponding one for uniformly distributed data. Slight modifications have been proposed to the original argument, see for instance Ye (2007).
\nocite{tibshirani2001estimating}\nocite{yan2007determining}

Another point of view useful to determine the number of clusters comes from model-based clustering. The idea is to fit a Gaussian mixture model to data and identify clusters as mixture components. The number of clusters is usually obtained using the well-known BIC criterion. All the above mentioned strategies seem to perform well when data do come from a mixture model but can perform poorly when the situation is more confused and fuzzy.

For determining the number of clusters James and Sugar (2003) proposed  an information theoretic approach. They consider the transformed distortion curve $d_K^{-p/2}$, a kind of average Mahalanobis distance between data and the set of cluster centers $C$ as a function of $K$. Jumps in the associated plot allow to select sensible values for $K$ while the largest one can be the best choice for a mixture of $p$-dimensional distributions with common covariance. An asymptotic analysis (as $p$ goes to infinity) states that, when the number of clusters used is smaller than the correct number (when any), then the transformed distortion remains close to zero, before jumping suddenly and increasing linearly. 
Then, detecting a jump in the transformed distortion curve is equivalent to detecting the number of clusters $K$.
\nocite{james2003finding}

One last element to help on the determination of the number of clusters is the validation tools. Once a clustering has been achieved one can use diagnostic tools in order to asses the quality of the clustering. Usually, clustering analysis is not a linear procedure. The practitioner must try several solutions so it will iterate between steps three and four. The final quality of a clustering output will be assessed by means of diagnostic displays that will be presented later.
              \section{Numerical illustration}\label{sec:sim}

We study the empirical performance of our clustering strategy on two functional data sets. The first one has been simulated using functional-valued processes and the second one is issued from the electricity power demand in France. 

           \subsection{Simulated example}   
We simulated a functional data set structured in three ($K = 3$) clusters. Each cluster is generated by a different functional model. 
An element of the first cluster is a simple superposition of two sinus and a white noise: $f(x) = \sin( 5 \pi x / 1024 ) + \sin( 2 \pi x / 1024) + \epsilon $. For the second and third clusters, we use two functional autoregressive (FAR) processes (see Bosq (2000)) defined by
 
   $$f_n = \rho f_{n-1} + \epsilon_n$$

Here for each $n\in\N$ a functional valued random value $f_n$ is observed, $\rho$ is a linear bounded operator and $\epsilon_n$ is a functional valued strong white noise. The first FAR has a diagonal covariance structure, and the second a full covariance matrix. The result is 25 segments of time series of length $1024$ composing each cluster. 
\nocite{bosq2000linear}

\begin{figure}
\centering
\includegraphics[width= 0.9\textwidth, bb=0 0 921 416]{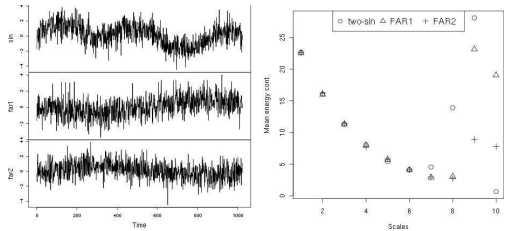}
  \caption{ On the left, some typical simulated trajectories of the sinus model 
	    (top panel), the \textsc{far1} model (middle), and the \textsc{far2} 
	    model (bottom). On the right, the mean scales' energy absolute 
	    contribution by cluster.}
  \label{fig:sim-traj}
\end{figure}

On the right of Figure \ref{fig:sim-traj} we plot one trajectory for each cluster. Note that the first model, dominated by a low frequency trend, is clearly distinguished from the two others whose differences are more intricate.

We apply to the sampled curves the DWT using a \textit{Symmlet 6} wavelet (see Nason (2008)) and we extract the absolute contribution (AC) of energy for each detail level $j = 1, \ldots, 10$. We calculate the mean AC by cluster. The results are plotted on the left of Figure \ref{fig:sim-traj}. Note how for a wide range of levels (from 1 to 6) there is no clear separation between cluster of the mean values of the AC. However, for the highest levels the extracted features show a better discrimination.
\nocite{nason2008wavelet}

To effectively detect which are the informative scales we use the Steinley and
Brusco's algorithm for variable selection for unsupervised learning. It needs as input the number of clusters so we perform the feature selection sequentially over a likely number of clusters ranging from two to twenty. We retain scales 8 to 10 which are associated with the lowest frequencies. We use the $k-$means algorithm (that we initialize many times, retaining the minimum within cluster distance solution) where the input data are the selected features and the number of clusters is the true one. We obtain so the predicted membership of each instance that we compare with the true ones. We repeat the process of generation of the three clusters, feature extraction, feature selection and clustering 100 times to eliminate the effect of the data generation. 

\begin{figure}
  \centering
  \includegraphics[width= 0.3\textwidth, height = 3cm]{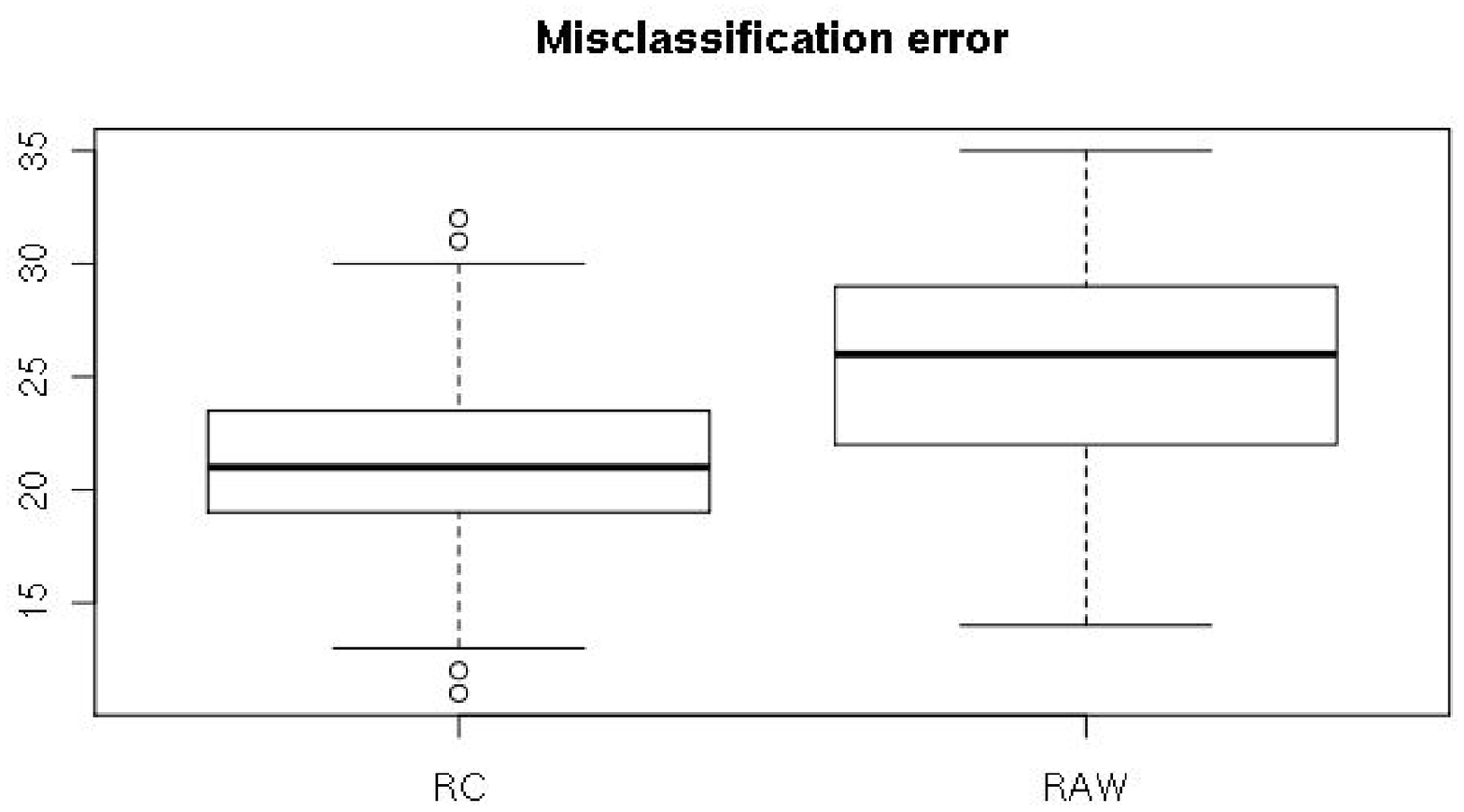} 
  \includegraphics[width= 0.3\textwidth, height = 3cm]{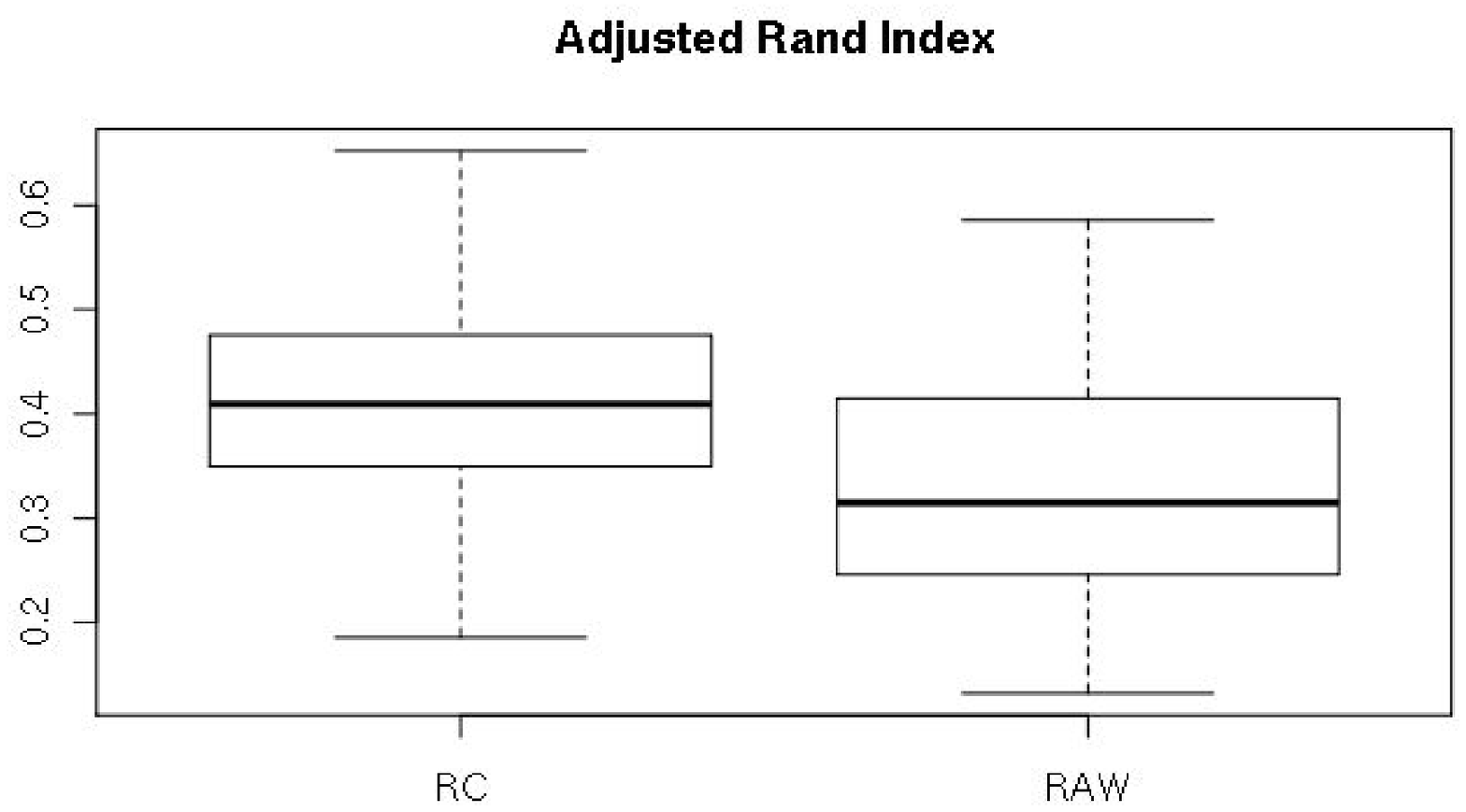}
  \caption{Boxplots of the misclassification error (left) and the Adjusted Rand Index for the 100 replicates of the simulated data set. Clustering using $k-$means on the extracted features (left) and on the raw curves (right).}
  \label{fig:simulation-resume}
\end{figure}

We then compare our clustering strategy with the one consisting in clustering the raw curves directly (i.e. clustering the observations as if they were vectors of dimension 1024). We will abbreviate RC the clustering that uses as extracted features the relative contributions of the energy and RAW the direct clustering of the curves. Table \ref{tab:resume} contains two quality indicators of the clustering while Figure \ref{fig:simulation-resume} presents the boxplots of the quality indicators across replications. 
   
\begin{table}
  \centering 
  \begin{tabular}{|ccc|} \hline 
   Clustering            & Feature Extraction & Raw curves    \\ \hline
    Abbreviation         & RC                 & RAW           \\
    Mean Global error    & 21.05 (4.125)      & 25.32 (4.834) \\ 
    Mean Rand Index      & 0.414 (0.092)      & 0.335 (0.109) \\ \hline 
  \end{tabular}
  \caption{Indicators of the clustering quality. Mean values over the 100 replicates with standard deviation between parenthesis.}\label{tab:resume}
\end{table}
  
We measure the global quality of the clustering by counting the number of observation misclassified and the Rand Index. This index measures the agreement between two clustering partitions: we will compare the clustering output and the real classes that we simulate for this example. The index is computed a ratio between the number of agreements between the partitions and the total number of agreements and disagreements. When corrected it takes values between -1 and 1. Values close to 0 show partitions that  are not correlated, while if it is close to 1 in absolute value the partitions  are highly correlated. The mean difference of the global error between RC and RAW clustering is significantly less than zero ($p-$value: $1e^{-10}$). In the same way, we obtain a significant greater value of the Rand Index for RC clustering ($p-$value: $3.7e^{-8}$).

\begin{figure}
  \centering
  \includegraphics[width= 0.45\textwidth, height = 3cm]{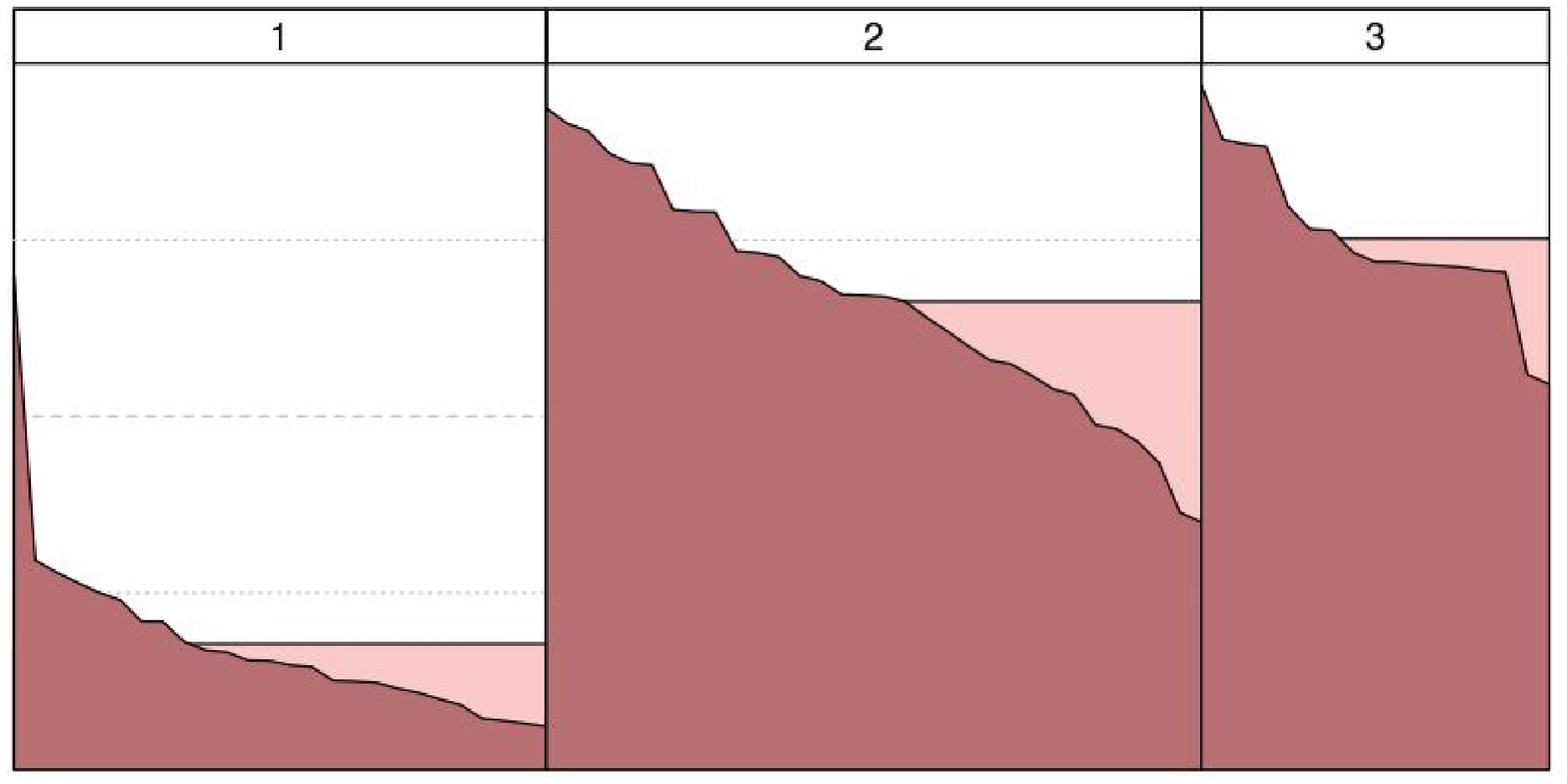}
  \includegraphics[width= 0.45\textwidth, height = 3cm]{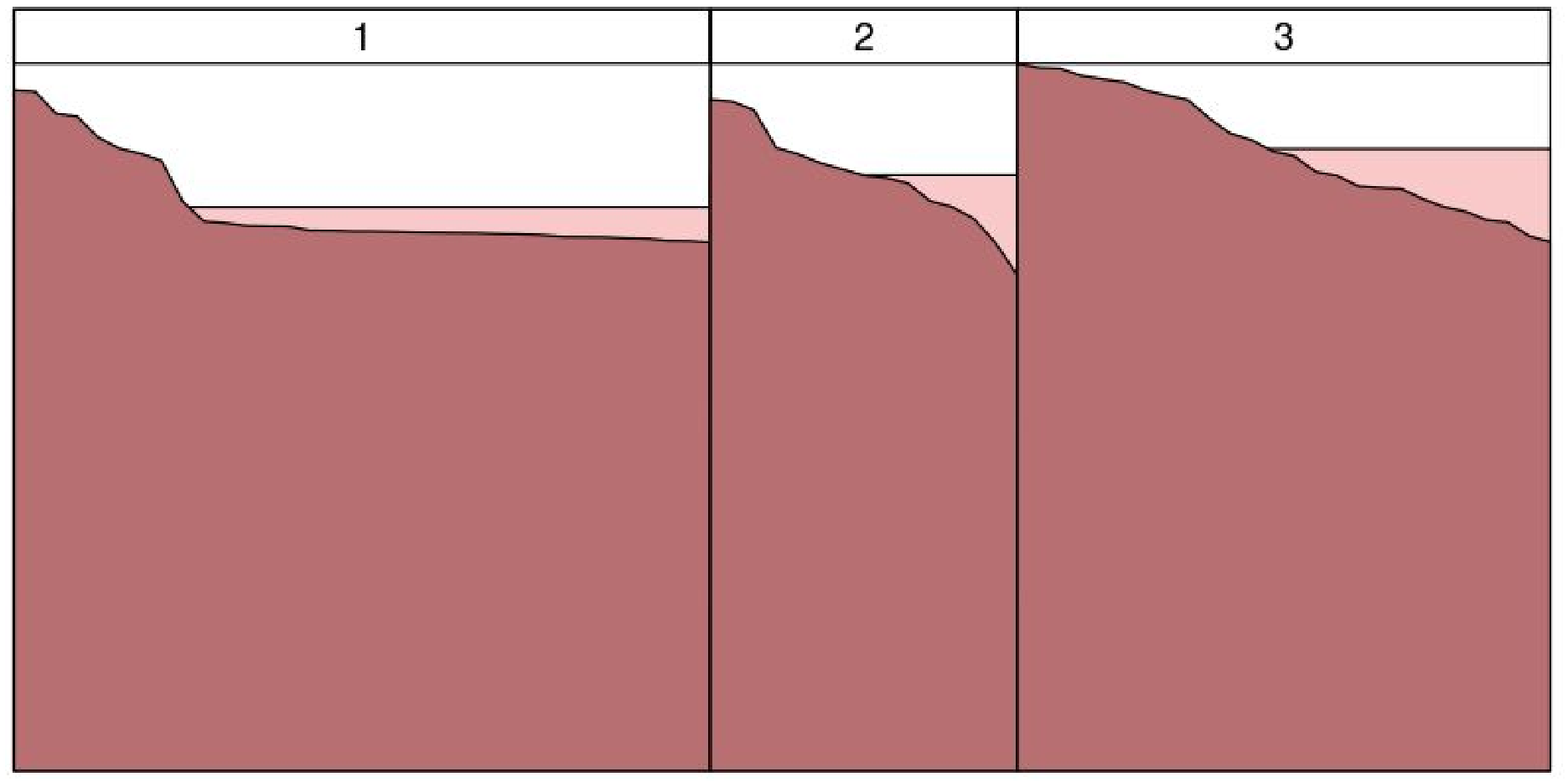}
  \caption{Shadow plots of resulting clustering for the simulated data set using $k-$means on the extracted feature (left) and on the raw curves (right).   Class 1 corresponds to the sinus model, Class 2 to the \textsc{far1} model and Class 3 to \textsc{far2} model.}
  \label{fig:simulation-shadow}
\end{figure}

Even with a lower global rate error, one may be interested in the quality of each cluster. We will help us by using the shadow plot (Leisch (2010)) to examine the intraclass quality. The shadow of an observation measures it distance to its centroid and to the second nearest centroid. If the observation is close to its centroid the shadow value is near 0, while if it is at the same distance between the first and second nearest centroids it gives values near to one. The shadow plot is a graphical representation of all shadow values arranged by cluster and sorted in decreasing order within the cluster. We represent the shadow plots of the resulting clustering of a randomly selected replicate in Figure \ref{fig:simulation-shadow}. The unequal width of each box is due to the different number of observations in each cluster. We see that globally the shadow values of the RC clustering (on the left) are lower than those of the RAW clustering (on the left) which shows that the former clustering provides more compact clusters than the latter one. Remark how the first cluster (the one that coincides with the sinus model) reveal a very compact cluster for the RC clustering. If the first cluster has such a good separation, most of the misclassification error must be made confusing clusters 2 and 3.
\nocite{leisch2009visualization}

Another useful diagnostic tool to assess the clustering quality of some high dimensional data is the neighborhood graph (Leisch (2006)). The main idea is to obtain some representation of the relative position of the clusters in some low dimensional space. As the projection to a plane may yield to a misinformation of how well two clusters are separated, the author proposes to combine it with a graph that reveals this information on the thickness of its edges. By this way, one constructs a graph where the nodes are the centroids and the thickness of the edges is proportional to the shadow mean value of the points belonging to the respective clusters of the relied centroids. To help the interpretation, one last element is added to the graphics: two convex hulls per cluster that will play an analogous role of the box plot for unidimensional data. The hulls formed by the thick lines correspond to the ``inner 50\%''. For each cluster $k = 1,\dots, K$, consider the distances from the centroid to each point of the cluster. Let be $m_k$ be the median distance of the cluster $k$. The ``inner 50\%'' of the cluster is defined as all the points with distance from the centroid is less or equal $m_k$. The second hull, the dashed one, contains all the points which distance from the centroid is less or equal $2.5 * m_k$.
\nocite{leisch2006toolbox}

\begin{figure}
  \centering
  \includegraphics[width= 0.9\textwidth]{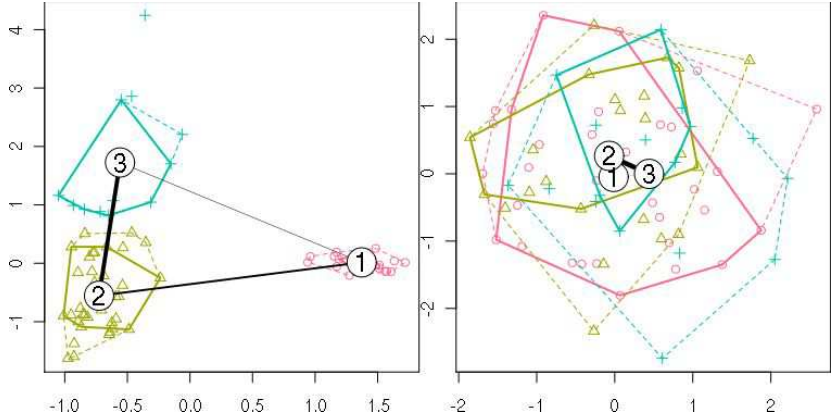}
  \caption{Neighborhood graph of resulting clustering for the simulated data set 
      using $k-$means on the extracted feature (left) and on the raw curves (right).}
  \label{fig:simulation-neig}
\end{figure}

The neighborhood graph of the RC and RAW clustering are shown on the left and right hand side respectively of Figure \ref{fig:simulation-neig}. For each clustering, the points are projected over the plane spanned by the two first principal directions of their respective spaces. The topology of both clusters is very different. While all the three convex hulls of the clusters on the RAW clustering overlaps in the principal plane, the separation of the clusters on the RC clustering is quite good. Thanks to the feature extraction, the cluster 1 (formed almost only by the sinus model observations) is completely separated from the rest of the observations on the principal plane. As we could anticipate in the shadow plots the separation of clusters two and three is less obvious.

\subsection{Electricity power demand data}

\begin{figure}
\centering
\includegraphics[width= 0.9\textwidth]{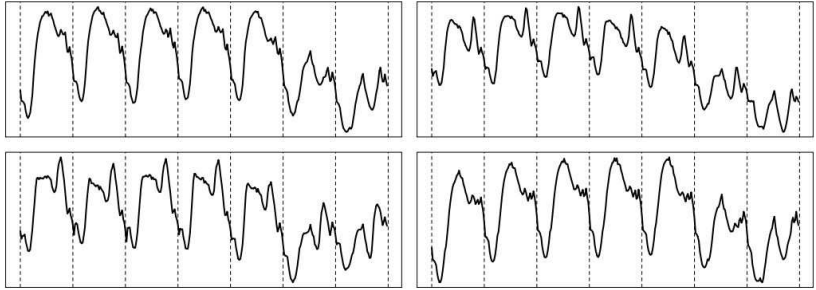}
\caption{Four weeks of the French electricity power demand: autumn (top left), winter (bottom left), spring (top right) and summer (bottom right). Divisions on the week trajectories mark the daily load curves. Weeks start on Mondays.}
\label{fig:conso-shapes}
\end{figure}

Our aim here is to discover clusters of daily load curves for a whole year. Before we start let us state some well known facts for \textsc{edf}'s experts from this data set. We will help us by using Figure \ref{fig:conso-shapes} where a sample of one week per season is presented. One important issue is the highly dependence on the meteorological conditions that is translated by a seasonal cycle. Also social and economic phenomenons (like holidays or the dichotomy between working-days and weekends) are present in the power demand. The structure of the weeks forms a weekly cycle where the profiles vary not only in mean level but also in the shape. It is also usual to found intraweekly differences, usually on the days next to weekends.

\begin{figure}
\centering
\includegraphics[width= 0.45\textwidth]{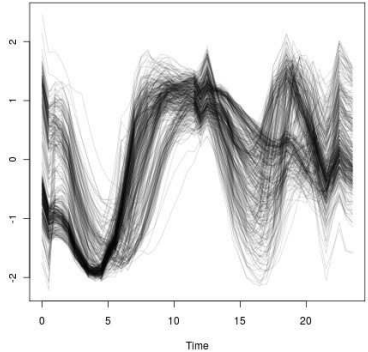}
\caption{Standardized data from French electrical power. Each curve corresponds to a daily profile of 2006.}
\label{fig:conso-traj}
\end{figure}

The centered profiles highlight the variation in shape mainly due to calendar structure (or to some special events like a few days when some industries are encouraged to  reduce their demand). But the variability in the curves is not only reduced to first and second order moments. The standardized version of daily profiles shows that higher order moments contribute also to the variability in the dynamic of the curves (see Figure \ref{fig:conso-traj}). The objective of this empirical evaluation is to discover groups that reflects this heterogeneity  to better understand the underlying structure. 

{\it Data preprocessing.} From the practical point of view we count have a discrete equidistant grid of 17520 ($=365 \hbox{x} 48$) time points of an underlying continuous process. After splitting the process as in (\ref{eq:X2Z}) with $\delta = 1$ day, the corresponding discrete versions of $(Z_n)$ are 48-length vectors $z_{i,J}, i=1,\ldots,365$. We use spline interpolation over each function in order to obtain $N = 64$ points  ($J = 6$) to be able to use Mallat's pyramidal algorithm for the DWT.

\subsubsection{Feature extraction and feature selection}
We proceed as before: for each $z_{i,6}, i=1,\ldots,365$ we compute the wavelet coefficients via the \textsc{dwt}. Then  we  calculate both the absolute and relative energy contributions of the scales $j = 1, \ldots, 6$ to the global
energy  (as in as in (\ref{eq:contribs})). We will called them AC and RC respectively. For the  RC we compute the logit transformations. We arrange the coefficients in two matrices of 365 rows and six  columns. 

For each data matrix the Steinley-Brusco's feature selection algorithm is used. As it needs as input the number of clusters we test it for a wide range of possibles number of clusters $k= 1, \ldots, K_{\hbox{max}}$  for some large positive $K_{\hbox{max}}$. For our application we used $K_{\hbox{max}} = 20$ ).

The algorithms returns which variables are significant for each $k$. The results of the algorithm show that

\begin{itemize}
  \item The significant scales for revealing the cluster structure are independent of the number of clusters used on the feature selection algorithm.
  \item As we could attend the significant scales are those associated with the mid-frequencies. In one hand, too large scales represents slow varying frequencies that are associated with the common day-night structure inherent to every day. In the other, too small scales captures very high frequency activity usually noise and thus no structure should be found neither.
  \item Finally, the scales that are significant may parametrize the cycle they represent. For the AC the scales that have been retained represent the cycles of 1.5, 3 and 6  hours. For the RC the scales retained are those associated to the cycles of 30 minutes, 1.5 and 3 hours. 
\end{itemize}

\subsubsection{Clustering results}
\begin{figure}
  \centering
  \includegraphics[width= 0.9\textwidth]{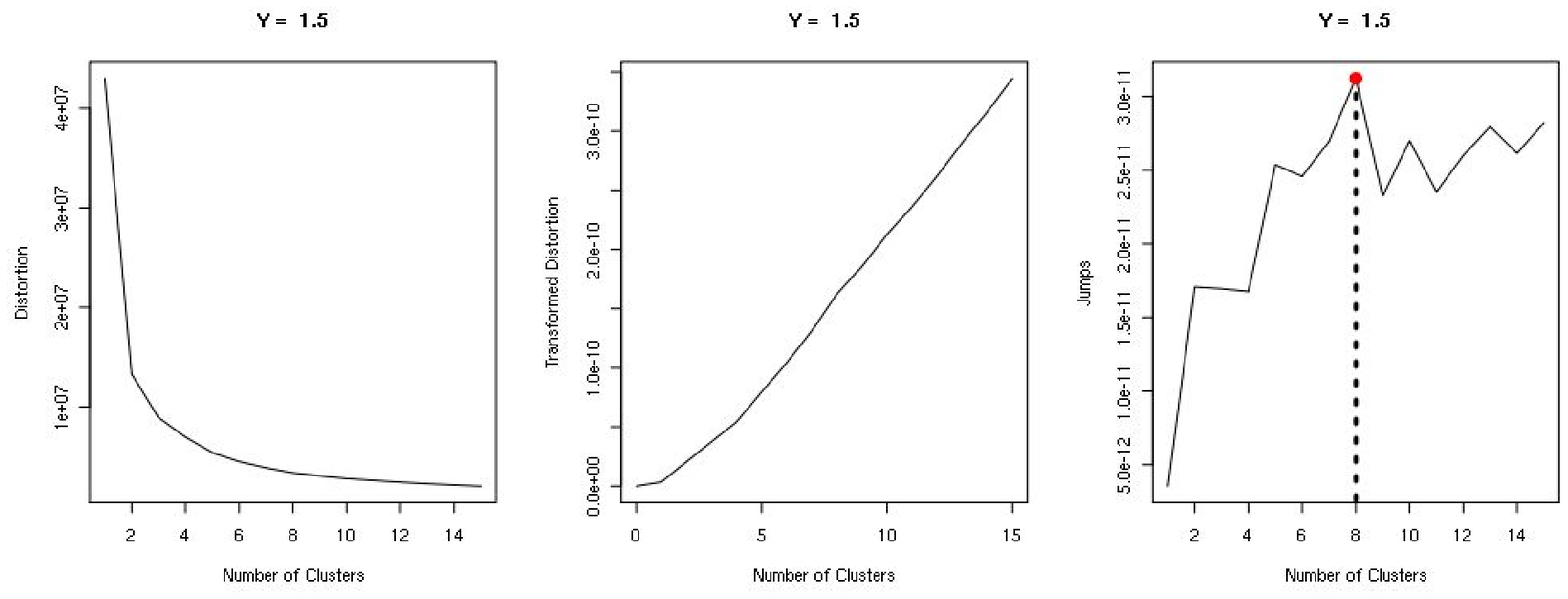}
  \includegraphics[width= 0.9\textwidth]{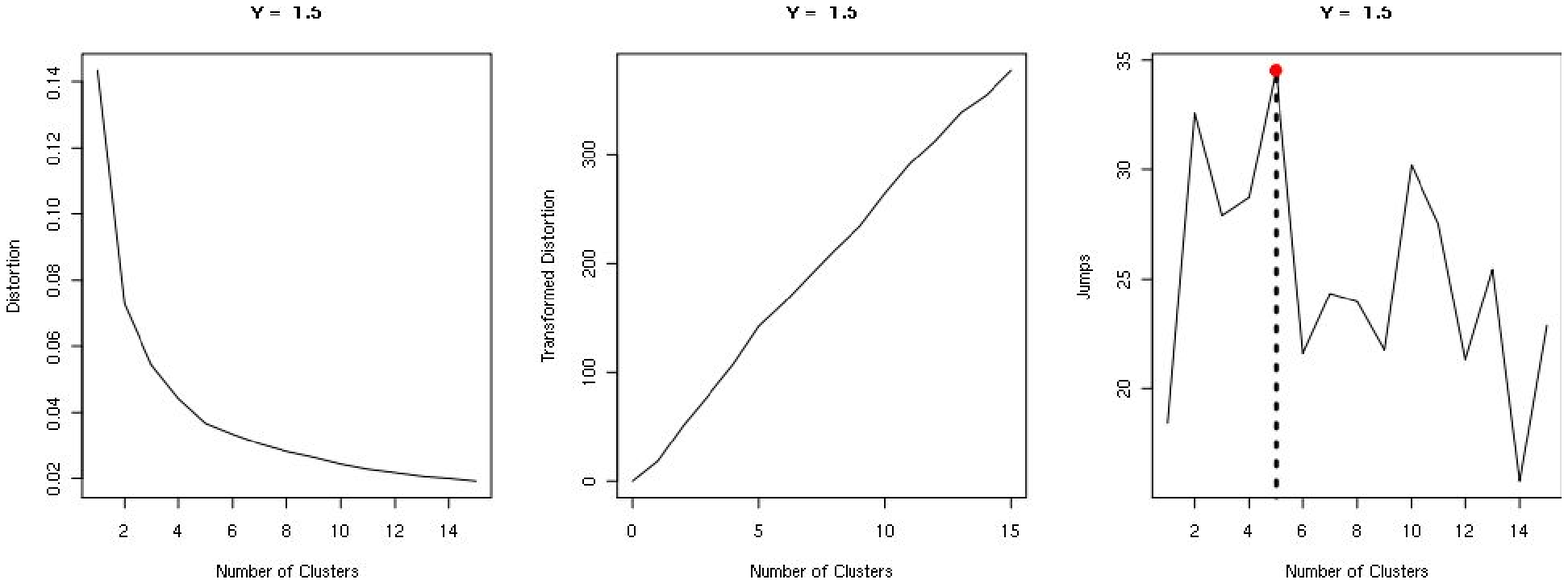}
  \caption{Detecting the number of clusters by feature extraction of the AC (top) and on the RC (bottom). From left to right we have the distortion curve, the transformed distortion curve and the first difference on the transformed distortion curve.}
  \label{fig:conso-jumps}
\end{figure}

The next step is to determine the number of clusters in data for both representations. We use the Sugar-James' own implementation of their algorithm to detect a jump in the transformed distortion curve. These curves are presented in Figure \ref{fig:conso-jumps}. 

For the AC and RC the number of detected clusters is 8 and 5 respectively. The difference in the number of  clusters is explained by the fact that while AC is vertical shift and scale invariant, the RC is only invariant by vertical shifts. Hence it founds less of variability in curves.

Then, we perform  the detection of groups using $k-$means algorithm. The input are the selected variables and the number of clusters detected by a significant jump in the distortion curve.

\begin{table}
\begin{tabular}{|l|c|c||l|c|c|} \hline
                & AC         & RC    &                           & AC   & RC\\ \hline 
Summer workdays & 1, 5, 7    & 1, 5  & Saturday                  & 3, 8 & 2 \\   
Cold weekdays   & 2, 4, 6    & 4     & Sundays and bank holiday  & 3, 8 & 3 \\  \hline
\end{tabular}
\centering
\caption{Resulting clustering for AC and RC variants.} \label{tab:conso-clutFE}
\end{table}

\begin{figure}
  \centering
  \includegraphics[width = 0.9\textwidth]{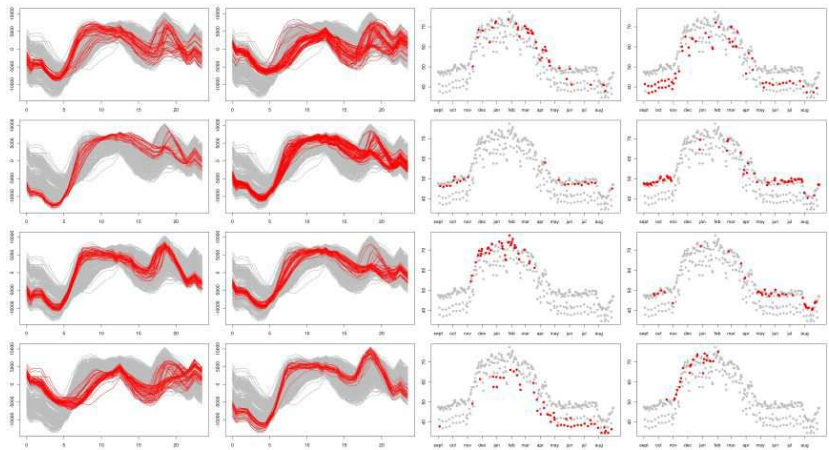}
  \caption{Curves membership (left) and calendar positioning (right) of the clustering using ACR feature extraction.} 
  \label{fig:conso-AC-calendar}
\end{figure}

\begin{figure}
  \centering
  \includegraphics[width = 0.9\textwidth]{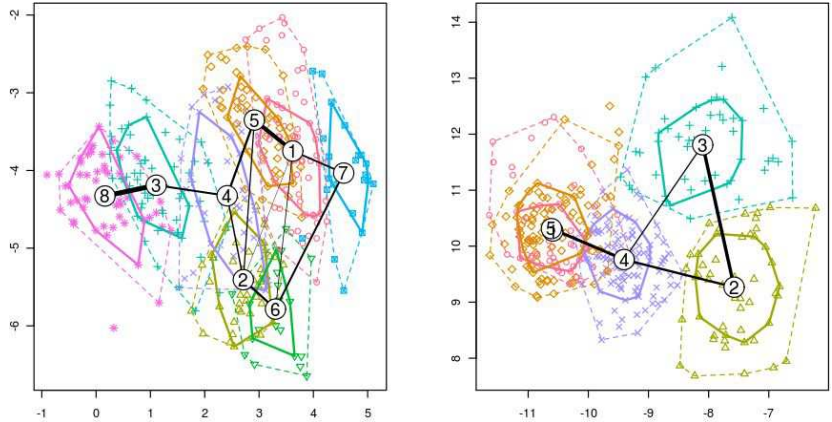}
  \caption{Neighborhood graphs of the two variants: AC (left) and RC (middle) for the \textsc{edf} data. }
  \label{fig:conso-AC}
\end{figure}

Let us sketch some interesting facts of the clustering results obtained from these both variants. First of all, we found for any of them what is well known in the electrical power demand environment: two well defined periods covering the summer and winter seasons. The clusterings also  found the dichotomy of working days and weekends. The clusters that correspond to these two season structure if resumed in Table~\ref{tab:conso-clutFE}. Figure \ref{fig:conso-AC-calendar} shows the clustering result founded using ACR. The figure is composed by two groups of 8 graphics (one for each cluster). The group on the left hand size shows the curves that were assigned to each cluster, while the one on the right hand size gives information about the position on the year of the instances of each cluster. The curves graphics (left hand side) have a gray shade formed by all the (standardized) daily curves is drawn on the background in order to help the visual comparison. In the calendar graphics (right hand side), each day in the year is represented by the mean average load in GW/h arranged chronologically. Again a gray shade is represents the set of all the days and only those corresponding to each cluster are highlighted. Finally, Figure \ref{fig:conso-AC} presents the neighborhood graphs of the resulting outputs.

A second issue is the difference between the AC and RC. Basically when passing from AC to RC one eliminates some of the heterogeneity in clusters. See for example how the three different clusters for cold workdays 6, 2 and 4 of AC (corresponding to early, medium and late winter respectively)  collapsed into one only cluster for the RC. This heterogeneity is consistent with the existence of two transition periods corresponding to the months of October-November and April-May. These transition days are no longer different from the winter workdays as soon as we do not authorize changes in scale (as mentioned before, RC is equivalent to standardized curves). So we can conclude that the transition days are different to the winter workdays on amplitude variation.

It is also interesting to see how the cluster are positioned in the space. The neighborhood graphs on the principal plane shows the summer clusters 1, 5 and 7 (corresponding to middle of the workdays, Fridays and summer break days and Mondays respectively) quite close between them, the winter clusters 2, 4 and 6 also closer between them, and Saturdays and Sundays clusters (3 and 8) slightly farther. There are also interesting links between the early winter cluster 6 and the late summer cluster 7 witnessed by an important edge that links them. 

Results are quite satisfactory but at this step one could ask itself whether the feature extraction suffices to well cluster a set of curves. Moreover, time aggregation on the computation of the energy contribution deletes all time location information. We will try in the next section to develop some dissimilarity measure that better exploit the power of the wavelet transform.

\section{Using the wavelet spectrums}\label{sec:cohe}

The success of any clustering algorithm depends on the adopted dissimilarity measure. Direct similarity measures such as $L_p$ norms match two functional objects in their original representations without explicit feature extraction. When $p=2$, this reduces to commonly used Euclidean distance. $L_p$ norms are straightforward and easy to compute. However, in many cases such as in shifting and scaling, the distance of two sequences cannot reflect the desired (dis)similarity between them.  Furthermore, $L_p$ distance has meaning only in the relative sense when used to measure (dis)similarity. 

In the previous section  we have  proposed instead the usage of the discrete wavelet transform of two curves of equal length to define a weighted normalized Euclidean like distance between them as a measure of their similarity. Indeed this was supported by the fact that the similarity of curves should be based on certain characteristics of the data rather than on the raw data itself by concentrating most of the energy in a small region of the scale-frequency domain. However the feature extraction procedure that we have used lost the location information. This is due to the aggregation on the time domain.

We may ask ourselves whether this loss is meaningful. To answer this question we present in this section two another direct and intuitive similarity measures that can be used for matching sequential patterns. 

The adopted similarity measures are based on the wavelet coherence between two time series (here considered as curves) and in a principal components analysis of the common covariance structure measured in terms of the cross wavelet transform. These concepts provide a way of analyzing local correlation or covariance of time series both in the time domain and in the frequency domain. In this, it fundamentally differs from Fourier coherence that relies upon the correlation of the two series in the frequency domain only.  In addition to locality, the continuous wavelet transform on which the wavelet coherence is based, possesses the very desirable ability of filtering the polynomial behavior to some predefined degree and therefore is invariant to vertical or scale shifts. Therefore, correct characterization of time series is possible, in particular in the presence of nonstationarities like global or local trends or biases. 

In what follows, we first recall some facts on the continuous wavelet transform and the concept of wavelet coherence between two time series. After that, we will use the maximum covariance analysis (MCA) for measuring the common time-frequency patterns and deduct a similarity measure between them.

\subsection{Continuous WT}

Although Fourier analysis is well suited to quantifying constant periodic components in a time-series, it is not able to characterize signals whose frequency content changes with time. On the other hand, a Fourier decomposition may determine all the spectral components embedded in a signal and does not provide any information about when they are present. To overcome this problem,  the wavelet transform decomposes a signal using functions (wavelets) that narrow when high-frequency features are present and widen on low-frequency structures (Daubechies (1992)). This decomposition yields a good localization in both time and frequency and is well suited for investigating the temporal evolution of aperiodic and transient signals. Indeed, wavelet analysis is the time-frequency decomposition with the optimal trade-off between time and frequency resolution (Mallat (1999)). 
\nocite{daubechies1992ten}\nocite{mallat1999wavelet}

Starting with a mother wavelet $\psi$, we consider the analyzing functions $\psi_{a,\tau}$ generated by simply scaling $\psi$ by $a>0$ and translating it by $\tau \in \mathbb{R}$

\begin{equation}\label{mothers}
   \psi_{a,\tau}(t) = \frac{1}{\sqrt{a}} \psi\left( \frac{t-\tau}{a}\right).
\end{equation}

The parameter $a$  is a scaling or dilation factor that controls the length of the wavelet (the factor $1/\sqrt{a}$ being introduced to guarantee preservation of the unit energy, $\| \psi_{a,\tau} \|_2 = 1$) and $\tau$ is a location parameter that indicates where the wavelet is centered. Scaling a wavelet simply  means stretching it (if $a > 1$), or compressing it (if $a<1$). Note that the choice of the wavelet function $\psi$ is not arbitrary. This function verifies also the moment condition $\int \psi(t) dt = 0$.

Given a function $z \in L_2(\mathbb{R})$, its continuous wavelet transform (CWT), with respect to the wavelet $\psi$, is a function $W_z(a,\tau)$ defined as

\begin{equation}\label{cwt}
W_z(a,\tau) = \int_{-\infty}^{\infty} z(t) \psi_{a,\tau}^*(t) dt,
\end{equation}

where `$*$' denotes the complex conjugate. The wavelet decomposition is a 
linear representation of the signal $z$ where the variance is preserved (Daubechies (1992)). Briefly, the continuous wavelet transform yields a redundant decomposition (the information extracted from a given scale slightly overlaps that extracted from neighbor scales) but it is generally more robust to noise as compared with other decomposition schemes (Poularikas (2009)). 
\nocite{daubechies1992ten}\nocite{poularikas2009transforms}

In some sense, the wavelet transform can be regarded as a generalization of the Fourier transform and by analogy  with spectral approaches, one can compute the local wavelet energy spectrum of $z$ defined by $S_z(a,\tau) = | W_z(a,\tau)|^2$ where $| \cdot |$ denotes the modulus. 

It is often desirable to quantify statistical relationships between two non-stationary signals. In Fourier analysis, the coherency is used to determine the association between two square-integrable signals $z$ and  $x$. The coherence function is a direct measure of the  correlation between the spectra of two time-series (Chatfield (1989)). To quantify the relationships between two non-stationary signals, the following quantities can be computed: the wavelet cross-spectrum and the wavelet coherence. 
\nocite{chatfield1989analysis}

The cross-wavelet transform is given by $\mathcal{W}_{z,x}(a,\tau) = W_z(a,\tau)W_x^*(a,\tau)$. As in the Fourier spectral approaches, the cross wavelet coherence can be defined as ratio  of the cross-wavelet spectrum to the product of the spectrum of each series, and can be thought of as the local correlation between two CWTs. Here, again, we follow Grinsted \textit{et al.} (2004) and define the wavelet coherence between two time series $z$ and $x$ as follows: 
\nocite{grinsted2004application} 

\begin{equation} \label{coherence}
R_{z,x}(a,\tau) = \frac{ |S(\mathcal{W}_{z,x}(a, \tau))| }{ |S(\mathcal{W}_{z,z}(a, \tau))|^{1/2} |S(\mathcal{W}_{x,x}(a, \tau))|^{1/2} },
\end{equation}

where $S$ denotes a smoothing operator in both time and scale (smoothing is  necessary since without that step, coherence is identically equal to 1). In Fourier analysis we overcome this problem by smoothing the cross-spectrum before normalization. For wavelet analysis and in particular with the Morlet wavelet, one can follow Torrence and Compo (1998), where the smoothing is achieved by a convolution in time and scale. The time convolution is done with a Gaussian window and the scale convolution is performed by a rectangular window (see Grinsted \textit{et al.} (2004) for details).
\nocite{torrence1998practical}
\nocite{grinsted2004application} 

We will now  explain our choice of wavelet coherence based distances motivated by the coefficient of determination $R^2$ from the linear regression framework.

\subsection{Extended coefficient of determination}
Consider a single linear regression between a response variable $Z$ and its regressor $X$. Given a set of sample data $(\mathbf{z}, \mathbf{x}) = (z_i, x_i)_{i=1,\dots, N}$ of length $N$, the model can be estimated in the least squares sense and a fitted line can be obtained. To measure how well does the adjusted regression line $\hat{\mathbf{z}}$ fits the data one should recall the definition of the coefficient of determination $R^2$: 

$$ R^2 = R^2 (\mathbf{z}, \mathbf{x}) = 
    \frac{\left[ \sum_{i=1}^{N} (z_i-\bar{z})({x}_i-\bar{x})\right]^2}
                {\sum_{i=1}^{N} (z_i-\bar{z})^2 \sum_{i=1}^{N} (x_i-\bar{x})^2}$$

where $\bar{z}$ indicates the average of $z$. Some simple calculations show that the coefficient of determination in simple linear regression is the square of Pearson's coefficient of linear correlation between $\mathbf{z}$ and $\mathbf{x}$. Therefore the coefficient $R^2$ measures the goodness of fit between the observed and the fitted values since it appears  as the cosine of the angle $\theta$ of the vectors $\mathbf{z}$ et $\hat {\mathbf{z}}$. It takes its values between 0 and 1 where the closer the value is to 1, the better the  regression line fits the data points. 

The coefficient of determination $R^2$ has some desirable properties as reflexivity ($R^2(\mathbf{z},\mathbf{z}) = 1$) or symmetry ($R^2(\mathbf{z}, \mathbf{x}) = R^2(\mathbf{x}, \mathbf{z})$). Thanks to these properties, $R^2$ appears to be a good similarity measure. Moreover $R^2$ has an intrinsic relation with the normalized Euclidean distance by mean-deviation. Indeed if $\nu(\mathbf{z})$ denotes the centered and standardized vector $\mathbf{x}$ and if $D_2(\nu(\mathbf{z}), \nu(\mathbf{x}))$ denotes the $L_2$ Euclidean distance between these two transformed vectors than it is easy to show that

  \begin{equation}\label{simdist}
    D_2(\nu(\mathbf{z}),\nu(\mathbf{x}) = \sqrt{2N(1-R^2(\mathbf{z}, \mathbf{x}))}.
  \end{equation}

Unfortunately,  $R^2$ is applicable only to 1-dimensional sequences, but in our application, the sequences we will consider will be multidimensional (indexed by the scales in the wavelet coherence vector). To match a pair of $J$-dimensional sequences $\mathbf{Z}$ and $\mathbf{Y}$, we will therefore use the extended coefficient of determination $ER^2$ defined as: 

 $$ER^2 = ER^2(\mathbf{Z}, \mathbf{X}) = 
    \frac{\left[ \sum_{j=1}^J  \sum_{i=1}^{N} (z_{ji}-\bar{z_j})  
                                              (x_{ji} - \bar{x_j}) \right]^2 }
                        {  \sum_{j=1}^J \sum_{i=1}^{N} (z_{ji}-\bar{z_j})^2 
                                        \sum_{i=1}^{N} (x_{ji}-\bar{x}_j)^2}.$$ 

where $\mathbf{z}_j$ and  $\mathbf{y}_j$  are the $j$th scale portions of $\mathbf{Z}$ and $\mathbf{Y}$. It is easy to see that $ER^2$  has properties similar to $R^2$.

\subsection{Scale-specific $ER^2$}
Direct assessment of relationships between time series at specific scales has remained a challenging problem.  Using the concept of wavelet coherence, we now show that traditional statistical measures, such as the extended multiple coefficient of determination analyzed in the previous subsection, can be adapted to address scale-specific questions in non stationary time series data. Scale-specific global relationships between two time series $z=(z(t); t\in\mathbb{R})$ and $x=(x(t); t\in\mathbb{R})$ can be quantified by computing first an averaged over time squared coherence

\begin{equation} \label{icoherence}
R_{z,x}^2 (a) =  \frac{ \int_{-\infty}^\infty |S(\mathcal{W}_{z,x}(a, \tau))|^2 d\tau}{ \int_{-\infty}^\infty |S(\mathcal{W}_{z,z}(a, \tau))| d\tau \int_{-\infty}^\infty |S(\mathcal{W}_{x,x}(a, \tau))| d\tau}.
\end{equation}

By definition (\ref{coherence})  the values of $R_{z,x}^2(a) $ are bounded by $ 0\leq R_{z,x}^2(a)  \leq 1$ exactly as it holds for the coefficient of determination $R^2$.  Moreover, the wavelet integrated over time squared coherency is equal to 1 when there is a strong  linear association at a particular scale between the two signals, and equal to 0 if $z$ and $x$ are non correlated. We may therefore use this  integrated squared wavelet coherence to define a scale-specific similarity measure $WER^2$ that looks like the familiar extended $ER^2$ based distance, say:

\begin{equation}\label{eq:wer}
  WER_{z,x}^2 = \frac{ 
 \int_0^\infty  \left( \int_{-\infty}^\infty |S(\mathcal{W}_{z,x}(a, \tau))|  d\tau \right)^2 da} { \int_0^\infty \left( \int_{-\infty}^\infty |S(\mathcal{W}_{z,z}(a, \tau))| d\tau \int_{-\infty}^\infty |S(\mathcal{W}_{x,x}(a, \tau))| d\tau\right) da}.
\end{equation}

Expression (\ref{eq:wer}) must be approximated in practice because we do not observe the continuous sample paths $(z(t), x(t))$. Instead, we have the samples  $\widetilde z = \{z(t_i)\}$ and $\widetilde x = \{x(t_i)\}$ for $i = 1, \ldots, N$. 
Hence, we must approximate the integral operation by summations over the $N$ time points. So in practice the CWT is computed only for $\tau = 1, \ldots, N$ and for an arbitrary set of scale values $a = \{a_j, j = 1,\ldots, J\}$. The smallest scale and the greatest scale are usually chosen as a power of two depending on the minimum detail resolution and the length of the time grid respectively. The rest of the values corresponds to a linear interpolation on a logarithmic  scale with base 2.

 These considerations yield in a $J \times N$ matrix $W_z$ whose $(k, j)-$th element  is

  $$ W_z (k, j) = \frac{1}{a_j} \sum_{i=0}^{N-1} z(t_i) \psi^{*} 
    \left( \frac{i-k}{a_j} \right) \qquad  k=0, \ldots, N-1, j=0, \ldots, J-1$$ 

The derived similarity measure, mimicking (\ref{simdist}) over the $J$ scales is then given by

 \begin{equation}\label{eq:dist-wer}
    d(z, x) = \sqrt{ J N \left(1 - \widetilde{WER}_{z, x}^2\right)} 
 \end{equation}

where $\widetilde{WER_{z, y}}$ is the analogous to (\ref{eq:wer}) calculated over a discrete grid where we have replaced integrals by summations over the scale set and the time-points set.

\subsection{MCA over the wavelet covariance}

We will explore a more sophisticated  alternative to measure non linear relationship between two non stationary time series. This approach uses a Maximum Covariance Analysis (MCA) over a localized covariance matrix based on the CWT. This way we can obtain time-frequency patterns that explains the principal covariation of the time series. We will introduce a way of quantifying the similarity of these patterns. Using MCA on the wavelet power spectrum of two series was originally proposed by Rouyer \textit{et al.} (2008). \nocite{rouyer2008analysing}. 

As before, consider two time series $z=(z(t), t\in\mathbb{R})$ and $z=(z(t), t\in\mathbb{R})$ and their CWT $W_z = W_z(a,\tau)$ and $W_x=W_x(a,\tau)$ computed from the sampled sample path $\tilde z$ and $\tilde x$.
We first define the time-frequency local covariance matrix by 

      $$Q_{zx} = W_z W_x^H $$

where $W_y^H$ is the conjugate transpose and $Q_{zx}$ is a $J\times J$  symmetric matrix with possibly complex values. Performing a SVD of $Q_{xy}$ gives the following decomposition

     $$ Q_{zx} = U \Gamma V^H$$

where the columns of $U$ and $V$ are the orthonormal singular vectors of $W_z$ and $W_x$ respectively, and $\Gamma$ is a diagonal matrix with the positive real numbers  $\lambda_1 \geq \ldots \geq \lambda_{J} \geq 0$ that we arrange in decreasing order. These numbers, known as the singular values of the decomposition give important information about $Q_{zx}$. For example the zero-norm gives the rank of the matrix. We have also that using the Frobenius norm $ \|Q_{zx}\|^2 = \| \Gamma \|^2 = \sum_{j=1}^{J} \lambda_j^2 $ so the total inertia of the covariance matrix is decomposed into the sum of squared singular values. By this way, the quantity $\lambda_j^2/\sum_j \lambda_j^2$ can be seen as the portion of explained covariance associated to the direction of the pair of the $j-$th singular vectors of $U$ and $V$ that we write $u_j$ and $v_j$ respectively.

We define also the $j-$th leading pattern as the projections of the CWT of $z$ and $x$ over their respective $j-$th singular vectors

 $$  L_z^j (t)  = u_j^H W_z  \quad \hbox{and} \quad    L_x^j (t)  = v_j^H W_x  $$

The leading patterns show how the wavelet scales evolve on time for the time series $z$ and $x$ in the orthogonal directions that maximize their common covariance. We can then decompose the CWT of $z$ and $x$ in terms of the singular vectors and the leading patterns obtaining

$$ W_z = U L_z \qquad \hbox{and} \qquad W_x = V L_x $$

where $L_z$ and $L_x$ are matrices that have the leading patterns of $z$ and $x$ respectively in their rows. As in PCA we chose the first $D$ directions for the smallest $D$ such that

 $$ \frac{\sum_{j=1}^{D} \lambda_j^2 }
         {\sum_{j=1}^{J} \lambda_j^2 } \geq \theta $$

where $\theta$ is a prefixed threshold. Thus, we obtain approximations of the CWT for both $z$ and $x$ using

 $$ W_z \approx \sum_{j=1}^{D} u_j L_z^j \quad \hbox{and} \quad
    W_x \approx \sum_{j=1}^{D} v_j L_x^j $$
    
where the informal notation $\approx$ is due to the truncation at the first $D$ direction for the expansions. This approximation guarantees a portion $\theta$ of the inertia of the covariance matrix $C$.

To compare the evolution on time of each pair of leading patterns we measure how dissimilar is their shape. For the $j$th pair of leading pattern, take the first derivative of the difference between them. The energy in this quantity is bigger if two leading patters presents very different evolutions. We finally measure this energy by taking the modulus 

   $$ d_j (z, x) = | \Delta(L_z^j - L_x^j) | $$

Finally, we aggregate all the significant directions using a weighted combinations with weights given by the explained square covariances

 $$ D(z, x) = \frac{\sum_{j=1}^{D} \lambda_j^2 d_j^2(z, x) }
                   {\sum_{j=1}^{D} \lambda_j^2 }$$

In the literature more complicated proposals for measuring parallelism between curves in similar contexts have been proposed. For example Keogh and Pazzani (1998) proposed as parallelism index to measure the angles  between the segments formed by each pair of consecutive points. 
Royuer \textit{et al.} (2008) use this index in the context of a MCA analysis on the wavelet power spectrum of two series. Aguiar and Soares (2009) use the MCA on a local covariation matrix based on the CWT with a complex wavelet. The parallelism index between the resulting complex valued leading patterns is measure as a complex angle of consecutive segments. Our proposal aims to measure the parallelism between complex valued vector avoiding to pass through the notion of complex angle.
\nocite{rouyer2008analysing}\nocite{keogh1998enhanced}\nocite{aguiar2009business}

\subsection{Clustering electricity power data through the wavelet spectrum.}

We test the proposed dissimilarities to cluster functional data on the the electricity power data. These dissimilarities impose a new challenge for the clustering algorithm: it is not clear why when using a distance other than the euclidean we should still calculate the centroids of the clusters by the average mean. We could for example argue that, as a consequence of the linearity of the WT the mean of elements is the mean element of a group. Otherwise, some deepness-based notion can be used to find some median-like element (e.g. see Cuevas \textit{et al.} (2006); Febrero \textit{et al.}(2008)). We will instead use the \texttt{cluster} R package to perform a partitioning around medoids (PAM). This technique admits a general dissimilarity matrix as input and is known to be more robust than $k-$means. The representative element of a group (medoid) is the element of the cluster that minimizes the dissimilarity to all the points of the cluster. \nocite{cuevas2006use, febrero2008outlier}

Each one of the daily segments is transformed by means of the \textsc{cwt} using the Morlet complex wavelet (see Mallat (1999)). We use the discrete scales set $\{2^j, j=1,\ldots,4\}$ and 8 octaves between dyadic scales.  The choice of this scales is made to filter the highest scales (above 12 hours) in order to eliminate the common low frequency pattern. The result is  a $41\times48$ complex matrix for each segment.
\nocite{mallat1999wavelet}

Then, a dissimilarity matrix is computed for the wavelet based extended $R^2$ dissimilarity (WER) and another for the one based on the maximum covariance analysis over the wavelet transform (MCA). We use the PAM clustering to obtain $k = 8$ clusters with each dissimilarity matrix. The number of clusters is chosen in order to be able to compare the clustering results with the AC clustering. 

\begin{table}
  \begin{tabular}{clcl} \hline \hline
    \textbf{A.} & Warming transition workdays & \textbf{E.} & Special days I   \\ 
    \textbf{B.} & Summer workdays             & \textbf{F.} & Hot Saturdays \\ 
    \textbf{C.} & Early winter workdays       & \textbf{G.} & Special days II  \\ 
    \textbf{D.} & Sundays   & \textbf{H.} & Late winter and cooling    \\
		&           &             & \hfill transition workdays \\ \hline
  \end{tabular}
  \centering 
  \caption {Labels for the MCA and WER clustering of the electrical power demand data.}
  \label{tab:conso-spectrum}
\end{table}

\begin{figure}
  \centering
   \includegraphics[width = 0.45\textwidth]{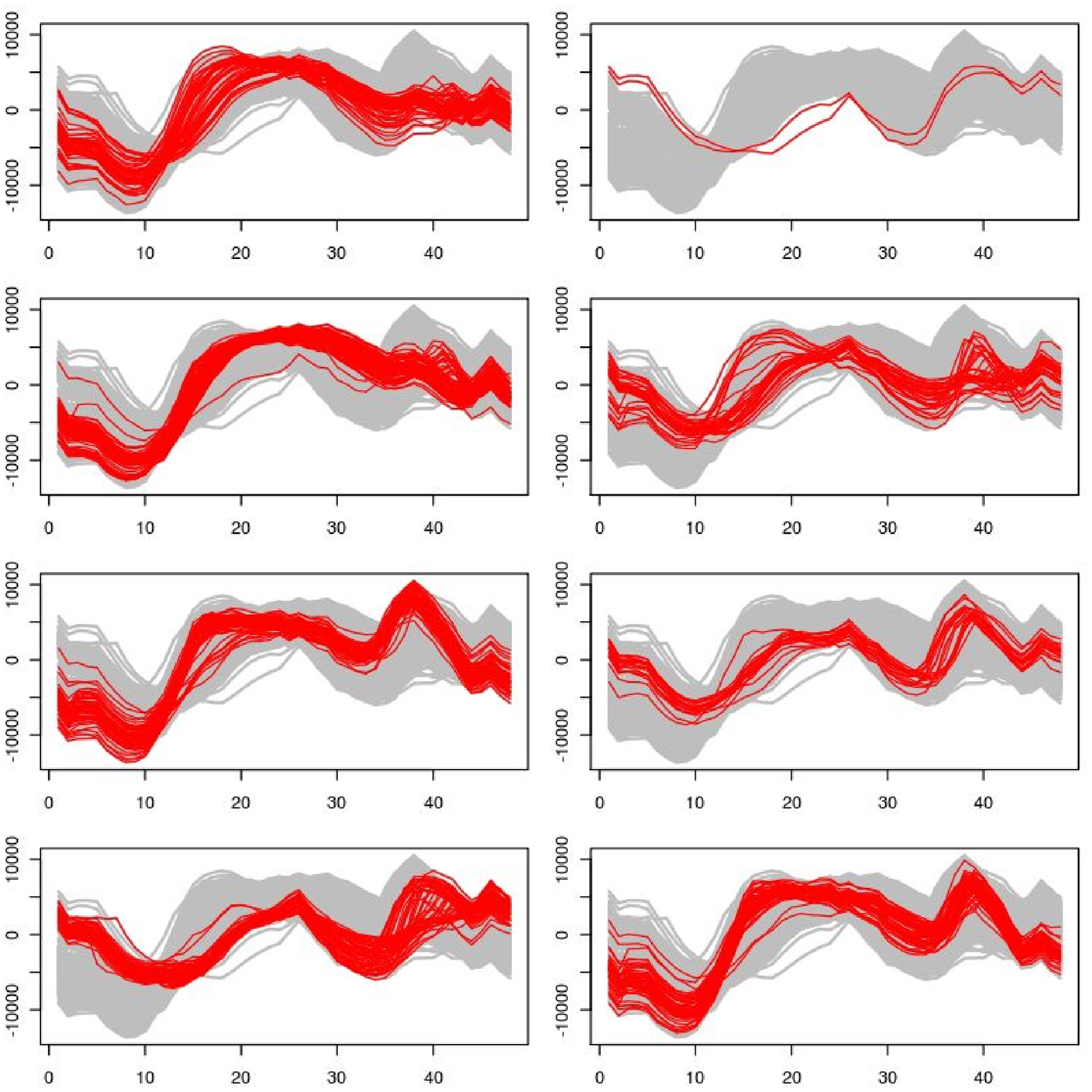}
   \includegraphics[width = 0.45\textwidth]{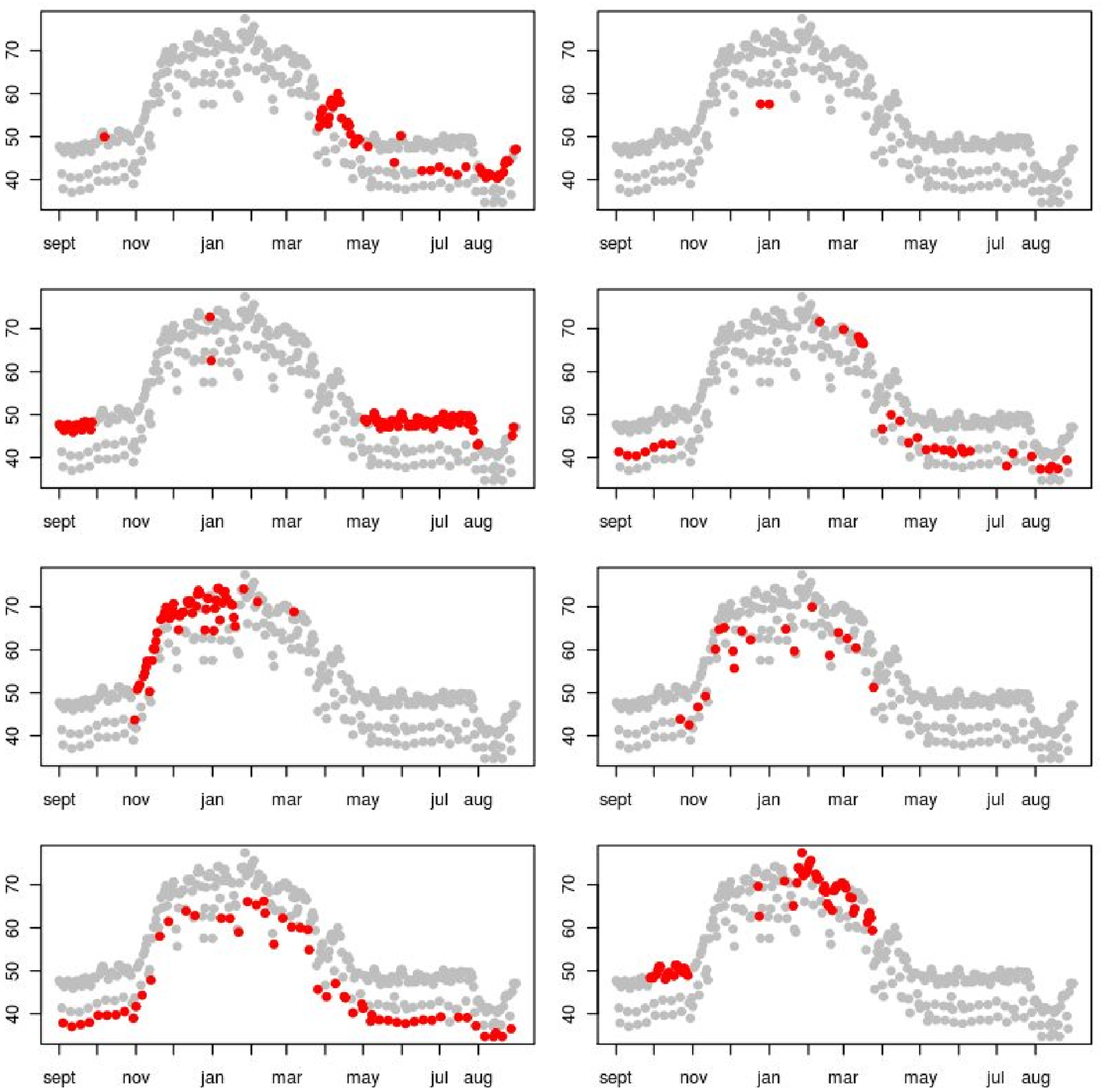} 
  \caption{Curves membership of the clustering using WER based dissimilarity (left) and the corresponding calendar positioning (right).}
  \label{fig:conso_clust_wer}
\end{figure}

\begin{figure}
  \centering
  \includegraphics[width = 0.45\textwidth]{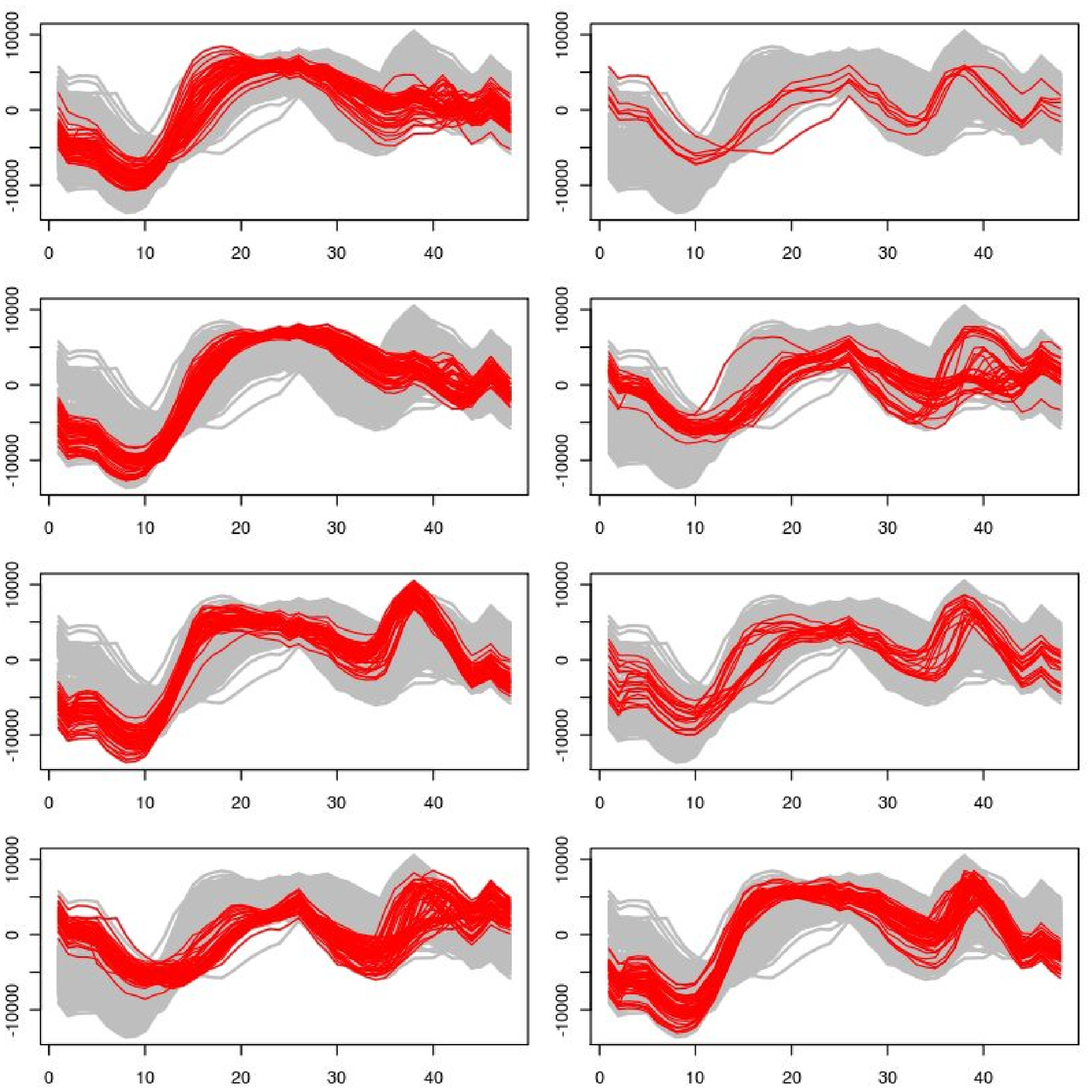}
  \includegraphics[width = 0.45\textwidth]{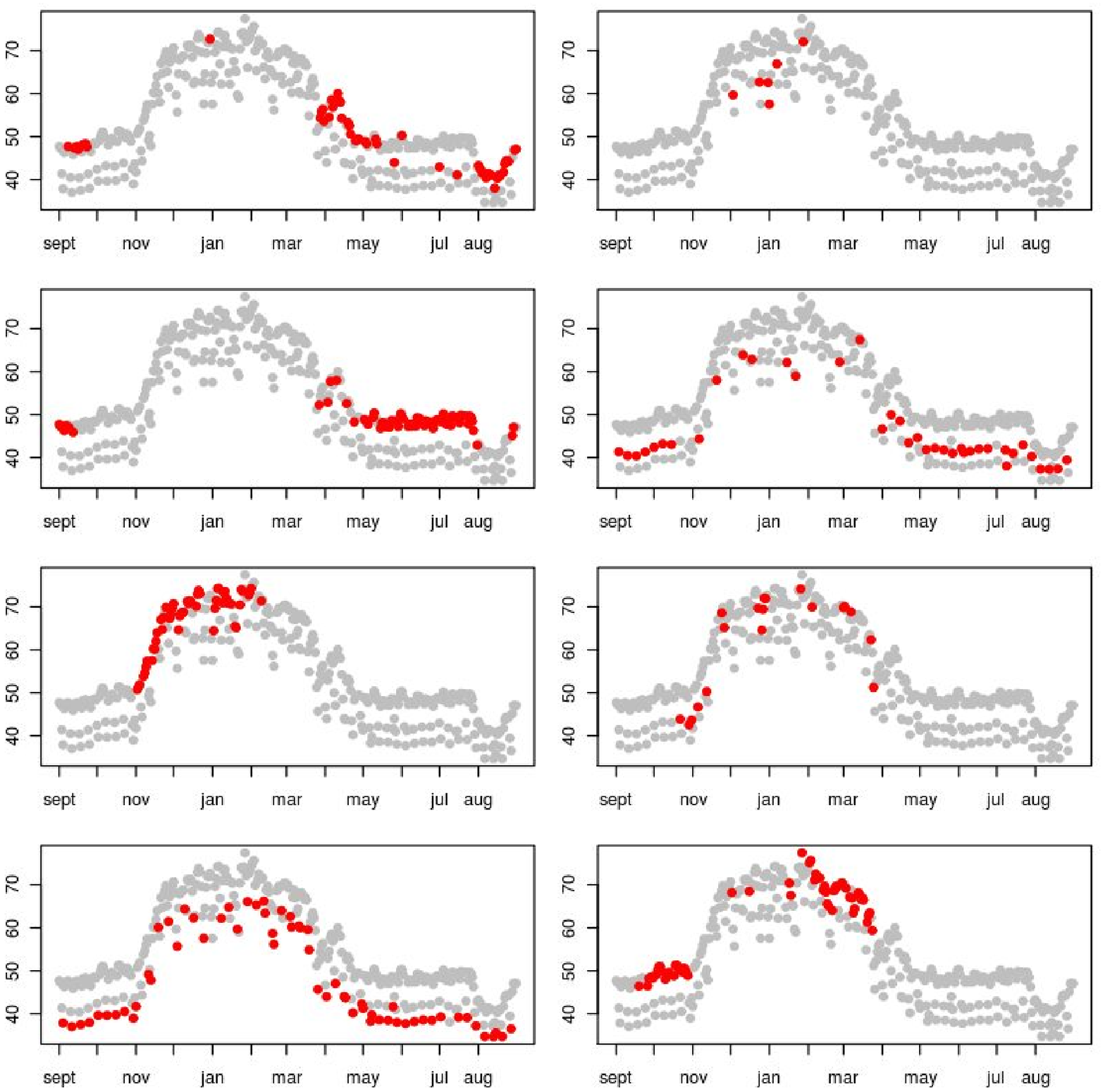}
  \caption{Curves membership of the clustering using MCA based dissimilarity (left) and
	    the corresponding calendar positioning (right).}
  \label{fig:conso_clust_mca}
\end{figure}

With the same display used for Figure \ref{fig:conso-AC-calendar}, Figures \ref{fig:conso_clust_wer} and \ref{fig:conso_clust_mca} show the clustering results founded using WER and MCA respectively. We compute the adjusted Rand index between the resulting partitions. While we obtain an Index of 0.26 and 0.32 between the AC clustering and WER and MCA respectively, the rand Index between WER and MCA is 0.57. This show that the resulting outputs of WER and MCA are quite different from that obtained with AC, and more consistent between them.  This is why we can use the same labels for the resulting clusters on both WER and MCA. Helping us with the calendar information we labeled the clusters as shown in Table~\ref{tab:conso-spectrum}. We leave two clusters with a generic names (special days I and II) to make a deeper analysis.

When we watch one year of daily curves, there is a large proportion of days where the dynamic of the demand is well known. This is the case of the clusters A, B, C, D, F and H where it is quite easy to find why they are together. Actually the graphics of the curves show that they are quite homogeneous groups with maybe some exceptions. However, important enhancement on the modeling on very few special tricky days can be made by founding this tricky days.

We focus our attention on the groups named ``Special days'' I and II (labels E and G respectively). These groups are not founded when using simpler clustering alternatives based on feature extraction. They are formed by only a few observations and they are not the same between the WER and MCA.

WER's cluster E only has two bank holiday (Christmas and New Year days) while for MCA clustering the cluster is formed by the eves of Christmas and New Year days and the New Year day as well as other three other cold weekend days. Cluster G has more elements (19 for both dissimilarities). For the WER dissimilarity it mainly formed by winter's Saturdays (16 out of 19). While for MCA the proportion of Saturdays is largely smaller (7 out 19) other winter tricky days are founded: the days of the one hour clock's shift due to the daylight saving time or the whole week between Christmas and New Year's day.  

\section{Concluding remarks}\label{sec:discussion}

The initial interest on clustering nonstationary functional time series was guided by the need for determining big structures of behavior of the daily power demand curves. The alternative would be to work directly over the vectors that constrains the sampled daily records using the $L_2$ metric. However, this turn out to be useless results in terms of clustering because one obtains groups that differ mainly in mean level and no information on the shape of the curves is recovered.

Our proposals for clustering functional data can be arranged into two types. While the first one is based on a wavelet-based feature extraction in order to use classical multidimensional clustering tools, the second way is to cluster using a dissimilarity measure between curves.
Both approaches are based on the wavelet transform. Our choice is guided by the interesting theoretical properties of wavelets. They are particularly fruitful in describing functional objects with localized structures.

The feature extracted clustering is extremely fast thanks to the fast implementation of the DWT and the $k-$means algorithm. We believe that the feature selection step is particularly useful.
On one hand, it eliminates non informative features that could induce an unsatisfactory clustering. On the other hand, it gives some adaptability to the technique with respect to the particular application and also gives interpretation  ability to the clustering technique.

In spite of the fact that the results of the first clustering strategy are useful in practice, more refined ones can be obtained using the dissimilarity based clustering proposals. On the daily load curves application we shown that very particular shapes of days can be founded with an extra computational burden. If for the WER dissimilarity this computational cost remains considerably low, the extra computational burden of the singular value decomposition for the MCA dissimilarity and the lack of meaningful different results from the WER dissimilarity drive us to think that it would not be useful in practice.
%
\bibliographystyle{plain}       
\bibliography{RR-7515}   

\begin{thebibliography}{10}

\bibitem{abraham2003unsupervised}
C.~Abraham, P.A. Cornillon, E.~Matzner-L{\o}ber, and N.~Molinari.
\newblock {Unsupervised curve clustering using B-splines}.
\newblock {\em Scandinavian Journal of Statistics}, 30(3):581--595, 2003.

\bibitem{aguiar2009business}
L.F. Aguiar and M.J. Soares.
\newblock {Business Cycle Synchronization Across the Euro-Area: a Wavelet
  Analysis}.
\newblock NIPE Working Papers 8/2009, NIPE - Universidade do Minho, 2009.

\bibitem{antoniadis2006functional}
A.~Antoniadis, E.~Paparoditis, and T.~Sapatinas.
\newblock {A functional wavelet-kernel approach for time series prediction}.
\newblock {\em Journal Royal Statisticial Society Series B Statistical
  Methodology}, 68(5):837, 2006.

\bibitem{antoniadis2003wavelet}
A.~Antoniadis and T.~Sapatinas.
\newblock {Wavelet methods for continuous-time prediction using Hilbert-valued
  autoregressive processes}.
\newblock {\em Journal of Multivariate Analysis}, 87(1):133--158, 2003.

\bibitem{besse1996approximation}
P.C. Besse and H.~Cardot.
\newblock {Approximation spline de la pr{\'e}vision d'un processus fonctionnel
  autor{\'e}gressif d'ordre 1}.
\newblock {\em Canadian Journal of Statistics}, 24(4):467--487, 1996.

\bibitem{bosq1991modelization}
D.~Bosq.
\newblock Modelization, nonparametric estimation and prediction for continuous
  time processes.
\newblock In George Roussas, editor, {\em Nonparametric functional estimation
  and related topics}, pages 509--529. NATO ASI Series, 1991.

\bibitem{bosq2000linear}
D.~Bosq.
\newblock {\em {Linear processes in function spaces: Theory and applications}}.
\newblock Springer-Verlag, New York, 2000.

\bibitem{chatfield1989analysis}
C.~Chatfield.
\newblock {\em {The Analysis of Time Series}}.
\newblock Ed. Chapman and Hall, 1989.

\bibitem{cuesta2007impartial}
J.A. Cuesta-Albertos and R.~Fraiman.
\newblock {Impartial trimmed k-means for functional data}.
\newblock {\em Computational Statistics \& Data Analysis}, 51(10):4864--4877,
  2007.

\bibitem{cuevas2006use}
A.~Cuevas, M.~Febrero, and R.~Fraiman.
\newblock {On the use of the bootstrap for estimating functions with functional
  data}.
\newblock {\em Computational statistics \& data analysis}, 51(2):1063--1074,
  2006.

\bibitem{daubechies1992ten}
I.~Daubechies.
\newblock {\em {Ten lectures on wavelets}}.
\newblock Society of Industrial Mathematics, 1992.

\bibitem{febrero2008outlier}
M.~Febrero, P.~Galeano, and W.~Gonz{\'a}lez-Manteiga.
\newblock {Outlier detection in functional data by depth measures, with
  application to identify abnormal NOx levels}.
\newblock {\em Environmetrics}, 19(4):331--345, 2008.

\bibitem{ferraty2006nonparametric}
F.~Ferraty and P.~Vieu.
\newblock {\em {Nonparametric functional data analysis: theory and practice}}.
\newblock Springer-Verlag, New York, 2006.

\bibitem{grinsted2004application}
A.~Grinsted, J.C. Moore, and S.~Jevrejeva.
\newblock {Application of the cross wavelet transform and wavelet coherence to
  geophysical time series}.
\newblock {\em Nonlinear Processes in Geophysics}, 11(5/6):561--566, 2004.

\bibitem{gurley2003first}
K.~Gurley, T.~Kijewski, and A.~Kareem.
\newblock {First-and higher-order correlation detection using wavelet
  transforms}.
\newblock {\em Journal of engineering mechanics}, 129(2):188, 2003.

\bibitem{james2003clustering}
G.M. James and C.A. Sugar.
\newblock {Clustering for Sparsely Sampled Functional Data.}
\newblock {\em Journal of the American Statistical Association},
  98(462):397--409, 2003.

\bibitem{james2003finding}
G.M. James and C.A. Sugar.
\newblock {Finding the Number of Clusters in a Dataset: An
  Information-Theoretic Approach.}
\newblock {\em Journal of the American Statistical Association},
  98(463):750--764, 2003.

\bibitem{keogh1998enhanced}
A.~Keogh and M.~Pazzani.
\newblock {An enhanced representation of time series which allows fast and
  accurate classification, clustering and relevance feedback}.
\newblock In {\em Proceedings of the 4th International Conference of Knowledge
  Discovery and Data Mining.} AAAI Press, 1998.

\bibitem{leisch2006toolbox}
F.~Leisch.
\newblock {A toolbox for k-centroids cluster analysis}.
\newblock {\em Computational Statistics and Data Analysis}, 51(2):526--544,
  2006.

\bibitem{leisch2009visualization}
F.~Leisch.
\newblock Neighborhood graphs, stripes and shadow plots for cluster
  visualization.
\newblock {\em Statistics and Computing}, 20(4):457--469, 2010.

\bibitem{luan2003}
Y.~Luan and H.~Li.
\newblock Clustering of time-course gene expression data using a mixed-effects
  model with b-splines.
\newblock {\em Bioinformatics}, (19):474---482, 2003.

\bibitem{mallat1989theory}
S.G. Mallat.
\newblock {A theory for multiresolution signal decomposition: The wavelet
  representation}.
\newblock {\em IEEE transaction on pattern analysis and machine intelligence},
  11(7):674--693, 1989.

\bibitem{mallat1999wavelet}
S.G. Mallat.
\newblock {\em {A wavelet tour of signal processing}}.
\newblock Academic Press, 1999.

\bibitem{nason2008wavelet}
G.~Nason.
\newblock {\em {Wavelet methods in statistics with R}}.
\newblock Springer, 2008.

\bibitem{percival2006wavelet}
D.B. Percival and A.T. Walden.
\newblock {\em {Wavelet methods for time series analysis}}.
\newblock Cambridge Univ Press, 2006.

\bibitem{poularikas2009transforms}
A.D. Poularikas.
\newblock {\em {Transforms and Applications Hanbook}}.
\newblock CRC Press, 2009.

\bibitem{pumo1992prediction}
B.~Pumo.
\newblock {\em {Estimation et pr{\'e}vision de processus autor{\'e}gressifs
  fonctionnels}}.
\newblock PhD thesis, University of Paris 6, 1992.

\bibitem{quiroga2004unsupervised}
R.Q. Quiroga, Z.~Nadasdy, and Y.~Ben-Shaul.
\newblock {Unsupervised spike detection and sorting with wavelets and
  superparamagnetic clustering}.
\newblock {\em Neural computation}, 16(8):1661--1687, 2004.

\bibitem{ramsay1991some}
J.O. Ramsay and C.J. Dalzell.
\newblock {Some tools for functional data analysis (with discussion)}.
\newblock {\em Journal of the Royal Statistical Society. Series B},
  53(3):539--572, 1991.

\bibitem{ramsay1997functional}
J.O. Ramsay and B.W. Silverman.
\newblock {\em {Functional data analysis}}.
\newblock Springer-Verlag, New Yok, 1997.

\bibitem{ramsay2002applied}
J.O. Ramsay and B.W. Silverman.
\newblock {\em {Applied functional data analysis}}.
\newblock Springer-Verlag, New York, 2002.

\bibitem{rouyer2008analysing}
T.~Rouyer, J.M. Fromentin, N.C. Stenseth, and B.~Cazelles.
\newblock {Analysing multiple time series and extending significance testing in
  wavelet analysis}.
\newblock {\em Marine Ecology Progress Series}, 359:11--23, 2008.

\bibitem{serban2004}
N.~Serban and L.~Wasserman.
\newblock {CATS}: clustering after transformation and smoothing.
\newblock {\em Journal of the American Statistical Association}, 100:990--999,
  2004.

\bibitem{steinley2008new}
D.~Steinley and M.J. Brusco.
\newblock {A new variable weighting and selection procedure for k-means cluster
  analysis.}
\newblock {\em Multivariate Behavioral Research}, 43(1):32, 2008.

\bibitem{steinley2008selection}
D.~Steinley and M.J. Brusco.
\newblock {Selection of variables in cluster analysis: An empirical comparison
  of eight procedures}.
\newblock {\em Psychometrika}, 73(1):125--144, 2008.

\bibitem{tarpey2007linear}
T.~Tarpey.
\newblock {Linear transformations and the k-means clustering algorithm:
  Applications to clustering curves}.
\newblock {\em The American Statistician}, 61(1):34, 2007.

\bibitem{tarpey2003clustering}
T.~Tarpey and K.K.J. Kinateder.
\newblock {Clustering functional data}.
\newblock {\em Journal of Classification}, 20(1):93--114, 2003.

\bibitem{tibshirani2001estimating}
R.~Tibshirani, G.~Walther, and T.~Hastie.
\newblock {Estimating the number of clusters in a data set via the gap
  statistic}.
\newblock {\em Journal of the Royal Statistical Society. Series B (Statistical
  Methodology)}, 63(2):411--423, 2001.

\bibitem{torrence1998practical}
C.~Torrence and G.P. Compo.
\newblock {A practical guide to wavelet analysis}.
\newblock {\em Bulletin of the American Meteorological Society}, 79(1):61--78,
  1998.

\bibitem{wang2008nonparametric}
H.~Wang, J.~Neill, and F.~Miller.
\newblock {Nonparametric clustering of functional data}.
\newblock {\em Statistics and its interface}, 1:47--62, 2008.

\bibitem{yan2007determining}
M.~Yan and K.~Ye.
\newblock {Determining the number of clusters using the weighted gap
  statistic}.
\newblock {\em Biometrics}, 63(4):1031--1037, 2007.

\end{thebibliography}
\end{document}